\documentclass{naturep}

\usepackage{amssymb}
\usepackage{amsmath}
\usepackage{graphicx}
\usepackage{algorithmicx}
\usepackage{algorithm}

\usepackage[noend]{algpseudocode}
\usepackage{bbm}
\usepackage{lineno}
\usepackage[margin=0.6in]{geometry}
\usepackage{ragged2e}
\usepackage{setspace}
\usepackage{longtable}
\usepackage{hyperref}
\usepackage{anyfontsize}
\usepackage{setspace}
\usepackage{float}
\usepackage{booktabs}
\usepackage{multirow}
\usepackage{xcolor}
\usepackage{blindtext}
\usepackage{tabularx}
\usepackage{caption}
\usepackage{subcaption}

\makeatletter
\let\saved@includegraphics\includegraphics
\AtBeginDocument{\let\includegraphics\saved@includegraphics}

\makeatother

\usepackage{enumitem, kantlipsum}

\title{\begin{flushleft}{\begin{spacing}{1}Unlocking adaptive digital pathology through dynamic feature learning\end{spacing}}\end{flushleft}}

\begin{document}

\maketitle
\begin{spacing}{1.8}
\vspace{-15mm}
\noindent Jiawen Li$^{1,\ddag}$, Tian Guan$^{1,\ddag}$, Qingxin Xia$^{6}$, Yizhi Wang$^{1}$, Xitong Ling$^{1}$, Jing Li$^{3}$, Qiang Huang$^{1,7}$, Zihan Wang$^{1,7}$, Zhiyuan Shen$^{1,7}$, Yifei Ma$^{2}$, Zimo Zhao$^{2}$, Zhe Lei$^{5}$, Tiandong Chen$^{6}$, Junbo Tan$^{1}$, Xueqian Wang$^{1}$, Xiu-Wu Bian$^{4,*}$, Zhe Wang$^{3,*}$, Lingchuan Guo$^{5,*}$, Chao He$^{2,*}$, Yonghong He$^{1,*}$
\end{spacing}
\vspace{-6mm}
\begin{spacing}{1.4}
\begin{affiliations}
 \item Tsinghua Shenzhen International Graduate School, Tsinghua University, Shenzhen, China
 \item Department of Engineering Science, University of Oxford, Oxford, UK
 \item State Key Laboratory of Holistic Integrative Management of Gastrointestinal Cancers, Department of Pathology, School of Basic Medicine and Xijing Hospital, Fourth Military Medical University, Xi’an, China
 \item Institute of Pathology and Southwest Cancer Center, Third Military Medical University, and Chongqing Advanced Pathology Research Institute, Jinfeng Laboratory, Chongqing, China
 \item Department of Pathology, the First Affiliated Hospital of Soochow University, and Institute of Clinical Pathology and Precision Medicine, Soochow University, Suzhou, China
 \item The Affiliated Cancer Hospital of Zhengzhou University and Henan Cancer Hospital, Zhengzhou, China
 \item Medical Optical Technology R\&D Center, Research Institute of Tsinghua, Pearl River Delta, Guangzhou, China
 \\$\boldsymbol{\ddag}$ These authors contributed equally
 \\$\boldsymbol{*}$ \textbf{Corresponding authors}: bianxiuwu@263.net (X.-W.B), zhwang@fmmu.edu.cn (Z.W.), \\szglc@hotmail.com (L.G.), chao.he@eng.ox.ac.uk (C.H.), heyh@sz.tsinghua.edu.cn (Y.H.)
\end{affiliations}
\end{spacing}
\newpage

\begin{spacing}{1.2}

\noindent\textbf{\Large{Abstract}}\\

\noindent \textbf{Foundation models have revolutionized the paradigm of digital pathology, as they leverage general-purpose features to emulate real-world pathological practices, enabling the quantitative analysis of critical histological patterns and the dissection of cancer-specific signals\cite{wang2024pathology,xu2024whole,vorontsov2024foundation,lu2024visual,chen2024towards,huang2023visual}. However, these static general features constrain the flexibility and pathological relevance in the ever-evolving needs of clinical applications, hindering the broad use of the current models\cite{zhang2024challenges,song2023artificial}. Here we introduce PathFiT, a dynamic feature learning method that can be effortlessly plugged into various pathology foundation models to unlock their adaptability. Meanwhile, PathFiT performs seamless implementation across diverse pathology applications regardless of downstream specificity. To validate PathFiT, we construct a digital pathology benchmark with over 20 terabytes of Internet and real-world data comprising 28 H\&E-stained tasks and 7 specialized imaging tasks including Masson’s Trichrome staining and immunofluorescence images. By applying PathFiT to the representative pathology foundation models, we demonstrate state-of-the-art performance on 34 out of 35 tasks, with significant improvements on 23 tasks and outperforming by 10.20\% on specialized imaging tasks. The superior performance and versatility of PathFiT open up new avenues in computational pathology.}
\end{spacing}
\newpage

\noindent\textbf{\large{Introduction}}

The advancements in computational pathology empower clinical applications through cancer diagnosis\cite{campanella2019clinical,lu2021ai,jiang2024transformer}, tumor subtyping\cite{coudray2018classification}, pathomics prediction\cite{chen2022pan,el2024whole}, and prognosis analysis\cite{lee2022derivation,volinsky2024prediction} from digitized tissue sections. Foundation models further accelerate the development of pathology-related AI tools\cite{huang2023visual,lu2024visual,chen2024towards,xu2024whole,wang2024pathology,vorontsov2024foundation,zhao2024foundation,kondepudi2024foundation}. By leveraging self-supervised learning on millions of tissue-contain image patches or regions of interest (ROIs) to capture universal clinical signals with histological patterns, those models provide general-purpose features for interpreting clinical gold standards\cite{song2023artificial}.

However, challenges still exist. Three main ones significantly hinder the practical application of pathology foundation models. First, the general features provided by fixed pretrain weights are not flexible for the diverse needs of real-world practice, hence these foundation models still cannot be widely used in clinical pathology. During the real-world procedure, pathological diagnosis exhibits significant biases, manifested in a large number of tumor types\cite{kundra2021oncotree}, complex morphological characteristics\cite{niazi2019digital,van2021deep}, and differences such as examination standards and data preprocessing methods across regions or medical institutions\cite{vaidya2024demographic}. These biases therefore make it difficult for general features to address specific tasks and contexts, limiting their effectiveness in clinical application. Second, foundation models still underperform in detecting fine-grained and rare diseases. To accurately diagnose these complex cases, it is necessary to capture subtle and specific pathological features. However, foundational models struggle to learn these signals from common histological datasets. For example, even foundation models trained on billions of image samples still struggle to accurately identify conditions like glioma and hepatobiliary carcinoma \cite{xu2024whole, vorontsov2024foundation}. Third, most of the data used for pretraining foundation models consist of H\&E-stained images. When these models are applied to tasks involving specialized imaging modalities such as Periodic Acid-Schiff (PAS) staining and immunofluorescence images, the general features they provide become less applicable\cite{zhang2024challenges,moor2023foundation}. For instance, the glomerular structure is complex and multifunctional, requiring Masson's Trichrome, PAS staining, or even immunofluorescence and transmission electron microscopy to highlight basement membrane thickening and assess lesion grade.

Here, we propose PathFiT, a dynamic feature learning method for unlocking adaptive pathology foundation models, aiming to provide a universal solution to these challenges. We notice that the typical use of foundation models is to extract static features as frozen encoders\cite{song2023artificial}. However, PathFiT can dynamically update foundation models based on clinical tasks to capture image features adaptively. The core of our method is: 1) to learn new knowledge without forgetting what has already been acquired, PathFiT freezes the original weights and introduces extra parameters\cite{hu2022lora} to update the foundation model \textbf{(Figure 1a,b)}, rather than updating the entire model weights. This re-embedding for general features allows learning dynamic signals while retaining the original representations; 2) to obtain new features without altering the modeling process, PathFiT parallelly integrates extra parameters into self-attention modules of foundation models to capture new dependencies between image tokens \textbf{(Figure 1c, Extended Data Figure 1)}. This plug-and-play operation of PathFiT can be applied to various pathological tasks while maintaining flexibility and stability.

We then construct a large-scale benchmark consisting of 35 clinically relevant tasks from both Internet and real-world data to show the adaptability of PathFiT for different clinical practice requirements. It covers a wide range of pathological data types, including H\&E-stained ROIs, biopsy and resection slides, and specialized pathology images (Masson's Trichrome, PAS, PASM, IHC-stained images, and immunofluorescence, transmission electron microscopy optical images). To validate PathFiT, we integrate it into the representative visual-language foundation model CONCH\cite{lu2024visual} and the visual foundation model UNI\cite{chen2024towards} \textbf{(Figure 1d)}. First, we demonstrate that PathFiT improves overall performance by 4.67\% compared to general feature learning, with significant improvements observed in 23 tasks. Second, overall 3.26\% and 5.91\% improvements on 9 fine-grained and 7 rare disease classification tasks demonstrate that the dynamic features of PathFiT effectively improve the ability to handle challenging tasks. Third, in specialized imaging tasks, PathFiT achieves a notable 10.20\% improvement, confirming that dynamic feature learning enhances foundation models with highly competitive capabilities in multimodal image analysis.

\begin{figure*}
\centering
\includegraphics[width=1\textwidth]{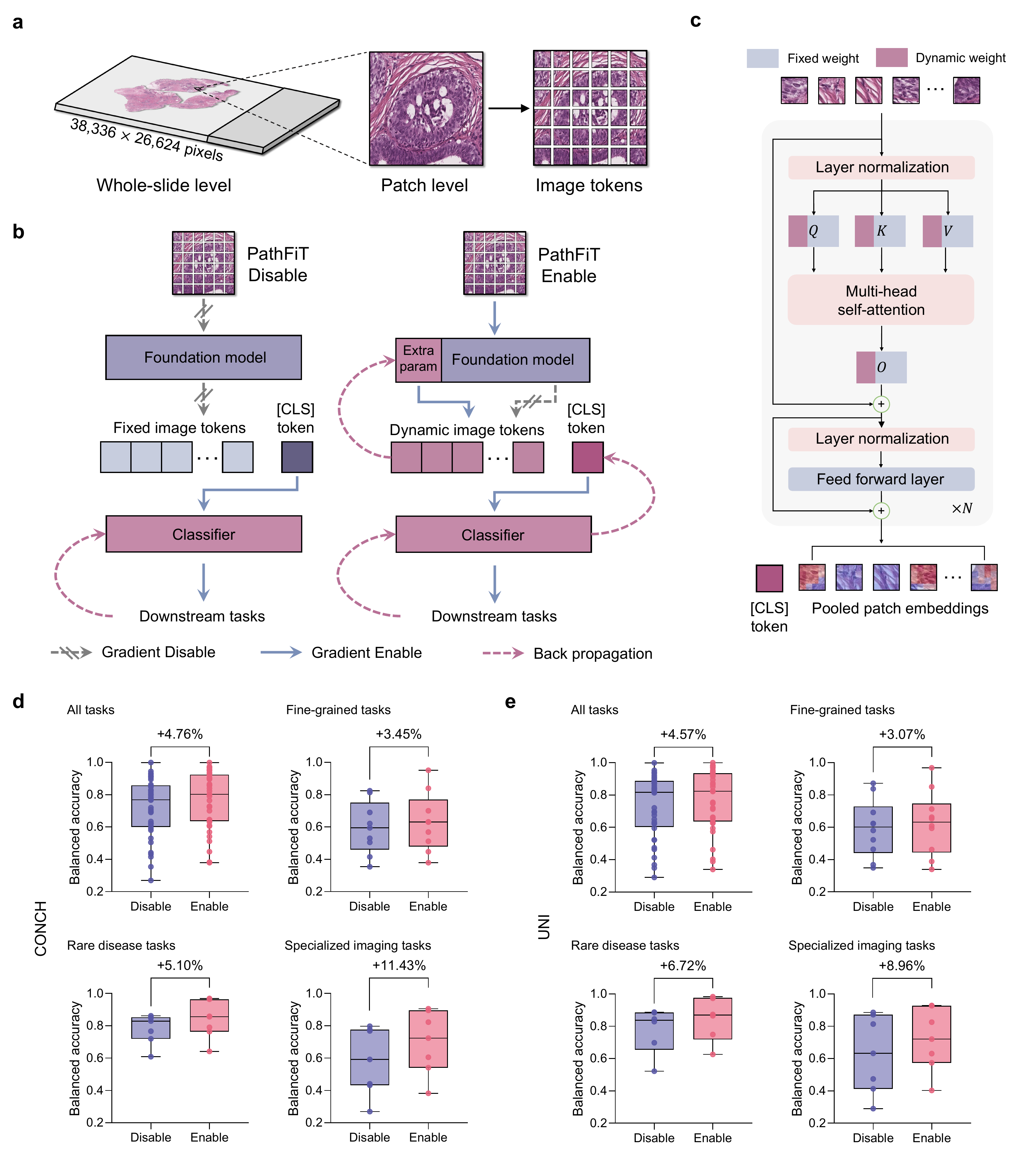}
\caption*{\textbf{Figure 1: Overview of PathFiT.} \textbf{a.} The typical paradigm in computational pathology is to use a series of tissue-contain patches as basic units, convert them into sequential image tokens, and feed them into transformer-based foundation models for forward modeling. \textbf{b.} The difference in downstream adaptation workflow between general feature learning and dynamic feature-based PathFiT. In the conventional process, only the parameters of the classifier layer are updated, while the weights within the foundation model remain unchanged. In contrast, PathFiT insets lightweight, trainable modules into the pretrained foundation model, enabling backpropagation to not only update the classifier but also dynamically adjust image features through the additional parameters to better adapt to downstream tasks. (Next page.)}
\end{figure*}
\begin{figure*}
\caption*{(Previous page.) \textbf{Figure 1: Overview of PathFiT.} \textbf{c.} PathFiT adds extra parameters in parallel to the linear layers within the self-attention of each transformer block. This design allows for dynamic adjustment of feature outputs while preserving the original model weights. \textbf{d.} PathFiT improves the performance of the visual-language foundation model CONCH on all tasks as well as fine-grained tasks, rare disease tasks, and specialized imaging tasks. \textbf{e.} PathFiT improves the performance of the visual foundation model UNI on all tasks as well as fine-grained tasks, rare disease tasks, and specialized imaging tasks.}
\end{figure*}

\noindent\textbf{\large{Results}}

\noindent\textbf{PathFiT improved resolution-agnostic ROI-level capabilities}

Histological diagnostics predominantly rely on H\&E-stained tissue sections as the foundation for analysis. ROIs within these sections often play a critical factor in uncovering disease mechanisms. By focusing on ROIs, AI can act as "second readers," complementing clinical workflows with precise and targeted insights. We assess the capabilities of PathFiT in ten ROI classification tasks. These include nine tasks from five subspecialties: 1) conventional subtyping (BACH)\cite{aresta2019bach} and fine-grained subtyping (BRACS)\cite{brancati2022bracs} in breast cancer, 2) precancer detection (MHIST)\cite{wei2021petri}, tissue classification (CRC-100K)\cite{kather2019predicting}, and microsatellite instability (MSI) status prediction (CRC-MSI)\cite{kather2020histological} in colorectal cancer, 3) tissue classification (KatherData) and MSI status prediction (KatherMS) in gastrointestinal cancers\cite{kather2019deep}, 4) tissue classification (OTA) in osteosarcoma\cite{arunachalam2019viable}, and 5) tissue classification (TolkachData) in esophageal cancer\cite{tolkach2023artificial}. Additionally, we conduct experiments on a large-scale pan-cancer classification task with 32 categories (TCGA)\cite{komura2022universal}. Due to the prevalent class imbalance in pathology tasks, we report balanced accuracy as the primary evaluation metric, as it provides a fair representation of model performance across all classes. The weighted F1 score and macro AUC are also reported to compare performance. \textbf{Extended Data Table 1-10} provide detailed experimental descriptions and specific results.

Our analysis demonstrated that PathFiT can consistently improve performance across all ten H\&E-stained ROI-level tasks for both foundation models. For CONCH, the overall AUC and balanced accuracy increased to 98.15\% and 91.08\%, with 1.86\% and 5.58\% improvement over disabling PathFiT. The balanced error rate decreased from 14.50\% to 8.92\%. Similarly, for UNI, AUC and balanced accuracy improved to 98.47\% and 92.46\%, with 1.44\% and 4.48\% increase over disabling PathFiT. The balanced error rate decreased from 11.81\% to 8.54\% \textbf{(Figure 2b-e)}. We noticed that some tasks approached performance limits, leading to diminishing marginal gains. We further analyzed the error reduction rate (ERR) for PathFiT across all tasks \textbf{(Extended Data Figure 2)}, providing a clear view of its improvement. Our experiments demonstrated that enabling PathFiT significantly reduced errors in both CONCH (overall ERR = 40.02\%) and UNI (overall ERR=39.29\%). For tasks nearing performance ceilings, such as CRC tissue classification (ERR=10.02\%, $p=0.02$ in CONCH; ERR=4.47\%, $p=0.41$ in UNI), GI tumor tissue classification (ERR=28.06\%, $p=0.16$ in CONCH; ERR=30.02\%, $p=4.00\times10^{-3}$ in UNI) and ESCA tissue classification (ERR=43.51\%, $p=0.02$ in CONCH; ERR=18.35\%, $p=0.44$ in UNI), PathFiT still achieved notable error reductions. In fine-grained or rare disease tasks such as BRCA fine-grained subtyping (ERR=5.98\%, $p=2.94\times10^{-3}$ in CONCH; ERR=7.95\%, $p=0.05$ in UNI) and CRC precancer detection (ERR=16.99\%, $p=0.01$ in CONCH; ERR=25.52\%, $p=2.51\times10^{-3}$ in UNI), PathFiT demonstrated consistent performance improvements. For the pan-cancer classification task, which demands a high level of feature representation, PathFiT enabled CONCH to achieve 95.06\% (+13.70\%, $p=1.86\times10^{-7}$) and UNI to achieve 96.74\% (+9.55\%, $p=2.47\times10^{-6}$).

Furthermore, we observed variations in the native resolutions of images across tasks. To evaluate this, we conducted experiments with four different resolutions on BRCA conventional subtyping, BRCA fine-grained subtyping, and OS tumor tissue classification tasks \textbf{(Figure 2g, Extended Data Table 38-43)}. Compared to general feature learning, enabling PathFiT consistently delivered superior performance such as 6.33\% ($p=6.07\times10^{-4}$), 7.09\% ($p=3.95\times10^{-3}$), 8.33\% ($p=1.42\times10^{-5}$) and 7.25\% ($p=1.87\times10^{-5}$) improvements in BRCA conventional subtyping across resolutions. We also noticed that PathFiT mitigated the performance degradation typically associated with increasing resolution, suggesting that its adaptability is resolution-agnostic. In addition, we visualized the features for qualitative analysis. On the BRCA conventional subtyping tasks, we used UMAP to reduce the dimensionality of ROI features to a 2D plane \textbf{(Figure 2h)}. The results showed that with PathFiT enabled, the cluster of each category became tighter, and the clusters of different categories became more distinct. This proved that dynamic learning adapted the general features to more task-specific embedding spaces. We also visualized the attention weights on the final layer of the foundation model to the corresponding image regions\cite{chen2022scaling} \textbf{(Figure 2i, Extended Data Figure 3)}. The generated heatmaps indicated that enabling PathFiT enhanced attention to diseased glands or cancer cell nuclei and reduced attention to irrelevant regions.

\noindent\textbf{PathFiT improved few-shot text prompt learning}

The scarcity of labeled images and the complexity of clinical tasks remain significant challenges in pathology image analysis\cite{nakagawa2023ai,perez2024guide}. Pathology foundation models not only need to identify morphological features in visual patterns accurately but also integrate closely with medical context and diagnostic knowledge. Visual models with single-modal may lack sufficient generalization due to their lack of cross-modal flexibility, particularly in leveraging natural language guidance\cite{zhou2022learning,zhou2022conditional}. Few-shot learning with text prompts offers dual benefits: 1) the training set requires only a small amount of data to achieve competitive performance\cite{shi2024vila,li2024diagnostic}, particularly for rare disease recognition\cite{li2024generalizable,chen2023dynamic}, 2) learning with a small number of image-text pairs helps to rapidly develop multimodal capabilities on visual foundation models, or reduce the time required to adjust prompt images or phrasing on vision-language foundation models. We evaluated PathFiT on pan-cancer classification, CRC tissue classification, and ESCA tissue classification tasks \textbf{(Figure 2f, Extended Table 44-49)}. The results demonstrated that for CONCH, while the 16-shot setting showed a slight performance drop (average 81.43\% vs 80.96\%, $p=0.13$), enabling PathFiT outperformed general feature learning across other few-shot settings, with average improvements of 8.39\% ($p=2.33\times10^{-5}$), 7.34\% ($p=5.28\times10^{-6}$), 5.53\% ($p=1.53\times10^{-4}$), and 3.21\% ($p=5.38\times10^{-4}$). For UNI, PathFiT achieved average improvements of 7.94\% ($p=1.70\times10^{-5}$), 5.65\% ($p=7.44\times10^{-4}$), and 5.03\% ($p=5.09\times10^{-4}$) in the first four shot settings, and showed a slight improvement in the 16-shot setting (average 0.44\%, $p=0.50$).

\begin{figure*}
\centering
\includegraphics[width=1\textwidth]{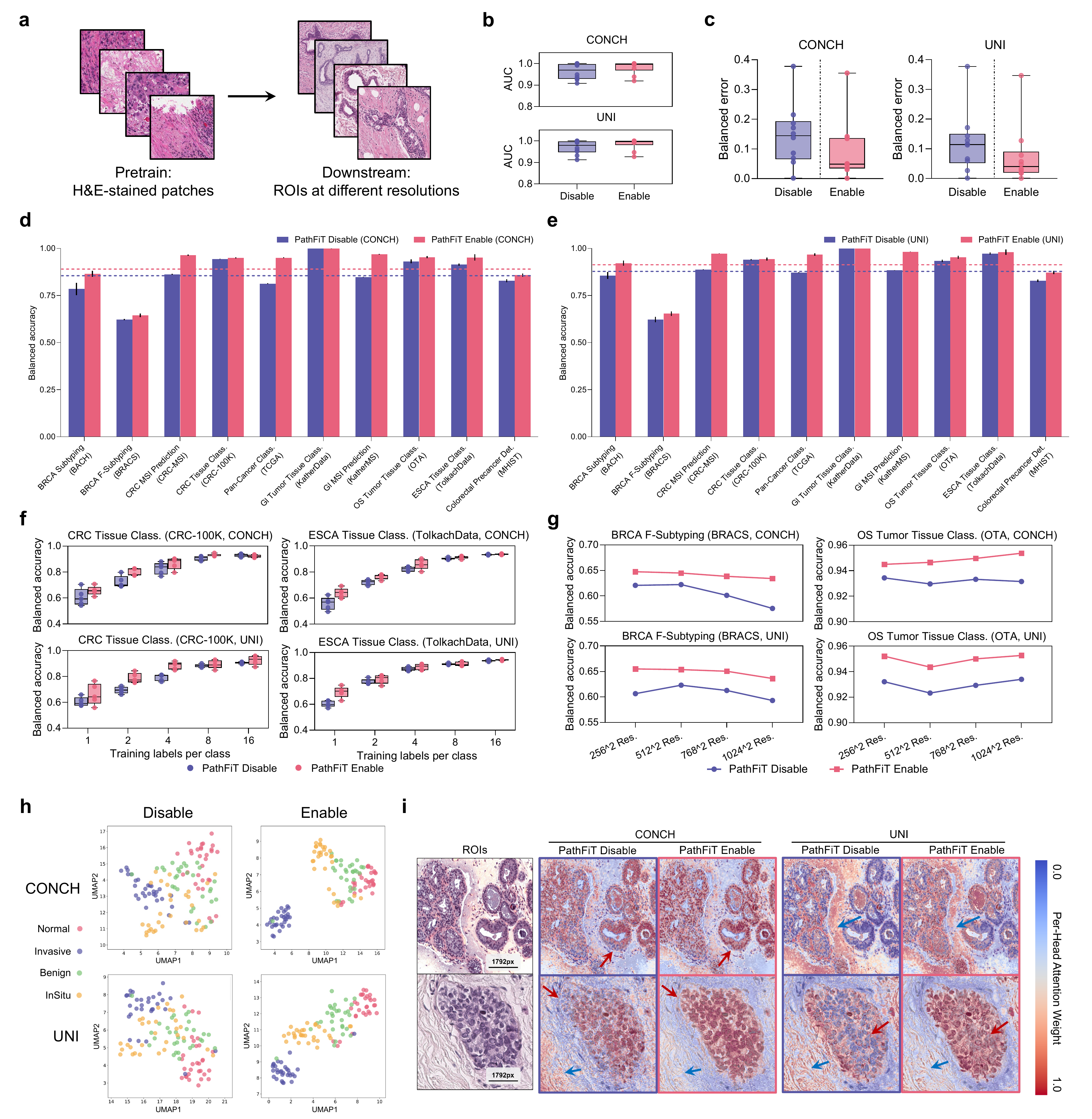}
\caption*{\textbf{Figure 2: ROI-level supervised classification.} \textbf{a.} By enabling PathFiT, foundation models pretrained on H\&E-stained image patches are adapted to ROI-level tasks at different resolutions. \textbf{b.} By enabling PathFiT, CONCH increased macro AUC from 96.29\% to 98.15\%, and UNI increased from 97.03\% to 98.47\%. \textbf{c.} By enabling PathFiT, CONCH decreased balanced error from 14.50\% to 8.92\%, and UNI decreased from 12.02\% to 7.54\%. \textbf{d,e.} Balanced accuracy comparison of CONCH and UNI across all ROI-level tasks between disabling and enabling PathFiT. \textbf{f.} Text prompt few-shot learning comparison between disabling and enabling PathFiT on CRC and ESCA tissue classification tasks. \textbf{g.} Comparison across different ROI resolutions between disabling and enabling PathFiT on BRCA fine-grained subtyping and OS tumor tissue classification. \textbf{h.} Visualization comparison of image embeddings between disabling and enabling PathFiT on the BRCA conventional subtyping task. \textbf{i.} Multi-head self-attention heatmap comparison with disabling and enabling PathFiT.}
\end{figure*}

\noindent\textbf{PathFiT improved pathology image segmentation}

The morphological features of nuclei and glands are crucial for building interpretable prognostic or diagnostic models\cite{huang2024pathologist}. To this day, segmenting nuclei or glands remains a challenging task in digital pathology. U-Net\cite{ronneberger2015u} has been one of the most widely used models for medical image segmentation due to its simplicity and lightweight structure, with numerous studies effectively validated on pathology images\cite{mahbod2024nuinsseg,kumar2017dataset}. To integrate pretrained foundation models seamlessly into a U-shaped architecture, we added a parallel branch on the encoder of U-Net to input images into the foundation model and the encoder simultaneously. The output of the foundation model is further fed into the decoder, and the encoder is connected to the decoder with skip connections. When PathFiT is enabled, the extra parameters in the foundation model and the encoder-decoder of the U-shape structure are updated together. When PathFiT is disabled, we ignore the extra parameters and only update the encoder-decoder. Unlike similar architectures such as TransUnet\cite{chen2024transunet}, our proposed framework enables plug-and-play functionality for the foundation model without requiring modifications to its internal structure, as is necessary with approaches like Mask2Former\cite{cheng2022masked}. We evaluated the framework on three tasks: epithelial cell segmentation with binary mask (SegPath)\cite{komura2023restaining}, colon gland segmentation (Warwick-QU)\cite{sirinukunwattana2017gland}, and multi-class semantic segmentation for colon nuclei identification (CoNIC)\cite{graham2024conic}. The dice score is used as the primary quantitative metric (Extended Data Table 11-13). Our results showed that enabling PathFiT generally outperformed fine-tuning with original weights. For CONCH, the average improvement is 0.58\% (-0.04\%, $p=0.84$ on SegPath; +0.63\%, $p=1.74\times10^{-3}$ on Warwick-QU; +1.17\%, $p=1.88\times10^{-3}$ on CoNIC). For UNI, the average improvement is 0.61\% (+0.22\%, $p=0.29$ on SegPath; +0.48\%, $p=0.01$ on Warwick-QU; +1.14\%, $p=8.85\times10^{-3}$ on CoNIC).

\noindent\textbf{PathFiT improved WSI classification}

\begin{figure*}
\centering
\includegraphics[width=1\textwidth]{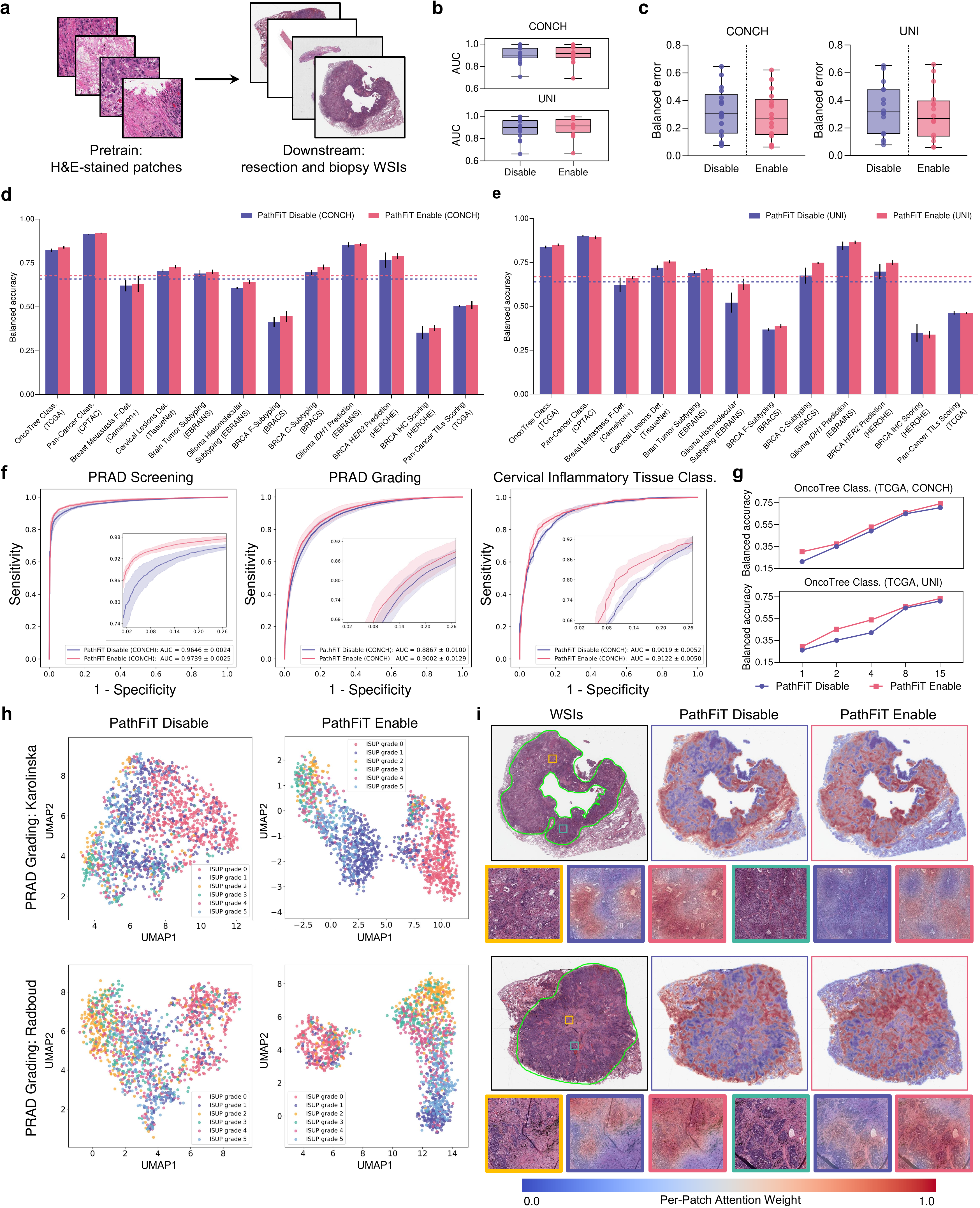}
\caption*{\textbf{Figure 3: Slide-level supervised classification.} Caption on next page.}
\end{figure*}
\begin{figure*}
  \caption*{(Previous page.) \textbf{Figure 3: Slide-level supervised classification.} \textbf{a.} By enabling PathFiT, foundation models pretrained on H\&E-stained image patches are adapted to resection and biopsy WSI tasks. \textbf{b.} By enabling PathFiT, CONCH increased macro AUC from 90.32\% to 90.58\%, and UNI increased from 88.53\% to 90.26\%. \textbf{c.} By enabling PathFiT, CONCH decreased balanced error from 31.68\% to 29.54\%, and UNI decreased from 33.60\% to 30.29\%. \textbf{d,e.} Balanced accuracy comparison of CONCH and UNI across all resection WSI tasks between disabling and enabling PathFiT. \textbf{f.} An average AUC of 97.39\%, 90.02\%, and 91.22\% was achieved for biopsy PRAD screening, PRAD grading, and cervical inflammatory tissue classification tasks with PathFiT enabled. \textbf{g.} Few-shot learning comparison between disabling and enabling PathFiT on TCGA OncoTree classification. \textbf{h.} Visualization comparison of image embeddings between disabling and enabling PathFiT on PRAD grading tasks. \textbf{i.} Attention weight heatmaps of the MIL aggregator between disabling and enabling PathFiT.}
\end{figure*}

Directly transferring foundation models to WSIs at full magnification involves converting the slide into extremely long sequences\cite{wang2023image}, which leads to an unrealistic increase in computational complexity. A conventional adaptation pipeline often extracts patch-level features from foreground tissues using foundation models, followed by training a multiple instance learning (MIL) structure\cite{lu2021data,li2021dual,li2024dynamic,yan2024shapley,tang2024feature} to aggregate these features and predict slide-level labels. Take ABMIL\cite{ilse2018attention} as an example: A gated attention mechanism is designed to generate attention scores and obtain slide-level representations by aggregating features of each patch. The key challenge here lies in adapting patch-level features to the global WSI feature space. PathFiT dynamically modifies a few or all patch features during adaptation, enabling online re-embedding to bridge the gap between upstream and downstream features. We evaluated PathFiT adaptation on ABMIL across fourteen gigapixel resection-level WSI tasks from eight cohorts (\textbf{Extended Data Table 14-25}), including OncoTree classification and pan-cancer tumor-immune lymphocyte (TILs) scoring from TCGA; pan-cancer classification from CPTAC; breast metastasis fine-grained detection\cite{ling2024towards} from Camelyon\cite{bejnordi2017diagnostic,litjens20181399,bandi2018detection}; cervical lesion detection from TissueNet\cite{lomenie2022can}; brain tumor subtyping, glioma histomolecular subtyping, and glioma \textit{IDH1} prediction from EBRAINS\cite{roetzer2022digital}; BRCA fine-grained and coarse-grained subtyping from BRACS\cite{brancati2022bracs}; and BRCA IHC scoring and \textit{HER2} prediction from HEROHE\cite{conde2021herohe}. Additionally, we evaluated three megapixel biopsy-level WSI tasks across 3 cohorts (\textbf{Extended Data Table 26-30}): PRAD screening and grading from PANDA (Radboud and Karolinska cohorts)\cite{bulten2022artificial}; and cervical inflammatory tissue classification from Xijing Hospital (XJH). For resection-level tasks, we randomly selected 64 patches in each iteration to update the extra modules due to the computational cost. For biopsy-level tasks, the number of tissue-containing patches is small in PANDA cohorts (maximum of 183), enabling all patch features to be updated in each iteration for both CONCH and UNI. For the XJH cohort, which contains more patches (from 2 to 4865), we only integrated PathFiT into CONCH to conduct experiments. 

Overall, enabling PathFiT increased CONCH and UNI to 90.58\% and 90.35\% in macro AUC \textbf{(Figure 3b)}, and decreased CONCH and UNI by 2.14\% and 3.38\% \textbf{(Figure 3c)}. Specifically, on resection-level WSI tasks, PathFiT delivered consistent improvements across almost all tasks, with average balanced accuracy gains of 1.82\% ($p=2.13\times10^{-4}$) for CONCH \textbf{(Figure 3d)} and 2.97\% ($p=2.07\times10^{-4}$) for UNI \textbf{(Figure 3e)}. For fine-grained and rare disease tasks such as those in the EBRAINS cohort, enabling PathFiT achieved superior performance with gains of 1.57\% ($p=0.05$) in brain tumor subtyping, 6.87\% ($p=0.03$) in glioma histomolecular subtyping, and 1.16\% ($p=0.06$) in glioma \textit{IDH1} prediction. On biopsy-level WSI tasks \textbf{(Figure 3f)}, enabling PathFiT improved balanced accuracy for CONCH by 1.69\% ($p=0.01$) and for UNI by 3.30\% ($p=4.07\times10^{-4}$) in PRAD screening compared to disabling PathFiT. The average AUC also reached 97.74\%, an increase of 1.82\% ($p=1.14\times10^{-5}$). PathFiT boosted balanced accuracy for CONCH on the cervical inflammatory tissue classification task by 4.42\% ($p=0.06$) and macro AUC by 1.03\% ($p=0.05$). In PRAD grading, PathFiT improved balanced accuracy by 3.34\% ($p=6.27\times10^{-5}$) for CONCH and by 5.36\% ($p=6.51\times10^{-4}$) for UNI. 

To investigate label efficiency on slide-level tasks with PathFiT enabled, we conducted few-shot learning experiments to evaluate 6 tasks (\textbf{Figure 3g}, \textbf{Extended Data Figure 4}). Overall, enabling PathFiT generally outperformed general adaptation. For CONCH, PathFiT showed superior performance in glioma histomolecular subtyping and BRCA \textit{HER2} prediction biomarker analysis, while achieving more stable performance improvements in BRCA coarse-grained subtyping as the number of shots increased. For UNI, although PathFiT did not meet expectations in the 4-shot setting of the \textit{HER2} prediction task, it achieved consistent improvements across other shot settings. Notably, PathFiT demonstrated greater performance gains for UNI than for CONCH in few-shot evaluations, highlighting its strong potential of PathFiT for large models with over 100 million parameters in real-world rare disease scenarios.

We used UMAP to visualize the slide embeddings between disabling and enabling PathFiT on PRAD grading tasks \textbf{(Figure 3h)}. The results demonstrated that re-embedding slide features using PathFiT can effectively distinguish the feature distribution of each slide, which is consistent with the results seen in ROI-level tasks. Furthermore, by using the CLAM tool\cite{lu2021data} to visualize the attention weights of the ABMIL in WSIs, changes in the weight distribution between disabling and enabling PathFiT were shown \textbf{(Figure 3i)}. We observed that PathFiT helped MIL aggregator focus on a broader range of lesion areas, and refined attention to local regions such as detailed diseased glands and cells. Also, the attention to non-diseased tissue regions has been reduced. \textbf{Extended Data Figure 5,6} provided more comparative heatmaps between disabling and enabling PathFiT.

\noindent\textbf{PathFiT improved specialized pathology imaging tasks}

\begin{figure*}
\centering
\includegraphics[width=1\textwidth]{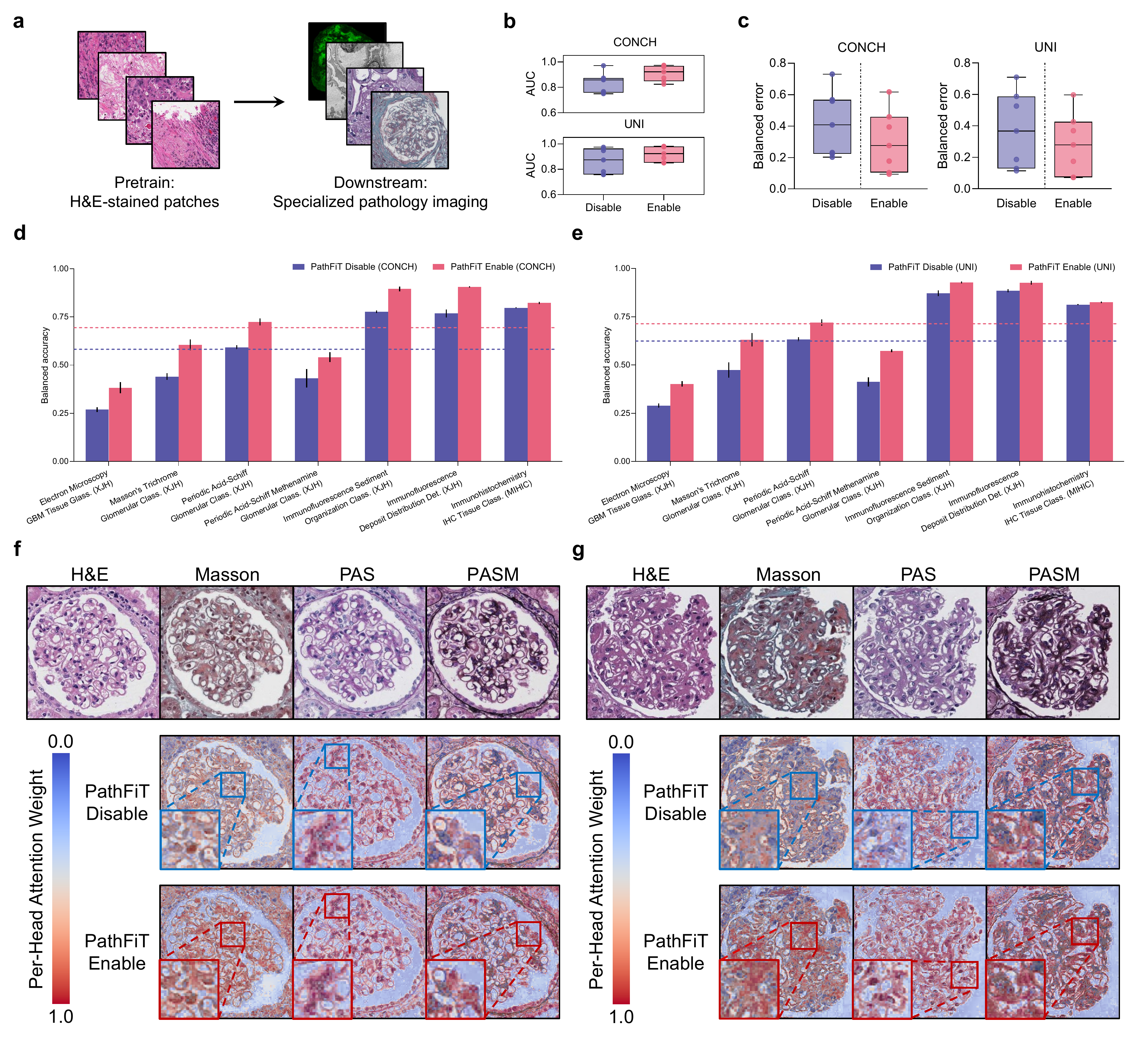}
\caption*{\textbf{Figure 4: Specialized pathology imaging classification.} \textbf{a.} By enabling PathFiT, foundation models pretrained on H\&E-stained image patches are adapted to specialized pathology imaging classification, such as Masson-stained, PASM-stained, transmission electron microscopy, and immunofluorescence images. \textbf{b.} By enabling PathFiT, CONCH increased macro AUC from 83.41\% to 91.07\%, and UNI increased from 86.46\% to 92.10\%. \textbf{c.} By enabling PathFiT, CONCH decreased balanced error from 41.77\% to 30.34\%, and UNI decreased from 37.44\% to 28.48\%. \textbf{d,e} Balanced accuracy comparison of CONCH and UNI across all specialized imaging tasks between disabling and enabling PathFiT. \textbf{f,g} Multi-head self-attention heatmap comparison on three special stains from the same glomerulus with disabling and enabling PathFiT.}
\end{figure*}

The capabilities of foundation models largely depend on the data alignment between downstream fine-tuning and upstream pretraining. Recent foundation models in computational pathology are predominantly pretrained on H\&E-stained images. However, many clinical practices rely on multimodal imaging data, making it difficult to consistently use general features extracted from foundation models for downstream learning. Some studies have attempted to incorporate immunohistochemistry and other stained images\cite{hua2024pathoduet} into pretraining databases, but the performance improvements remain limited.

To explore the ability of PathFiT on specialized staining and optic imaging modalities, we collected and constructed a specialized pathology imaging benchmark, which is the large-scale cross-domain pathology image database, consisting of 9656 images across 6 modalities from the Internet and the in-house Xijing Hospital. This database includes 4 types of special stains: Masson's Trichrome, Periodic Acid-Schiff (PAS), Periodic Acid-Schiff Methenamine (PASM), and immunohistochemistry (IHC), as well as two optical imaging modalities: immunofluorescence and transmission electron microscopy. We performed 7 clinically relevant tasks, including glomerular structure classification of transmission electron microscopy, Masson’s Trichrome glomerular classification, PAS glomerular classification, PASM glomerular classification, immunofluorescence sediment organization classification, immunofluorescence deposit distribution detection, and immunohistochemistry tissue classification.

Overall, compared to general feature learning, PathFiT improved the average macro AUC of CONCH and UNI by 7.66\% and 5.63\% \textbf{(Figure 4b)}, and reduced balanced error by 11.43\% and 8.97\% \textbf{(Figure 4c)}. Specifically, foundation models with PathFiT enabled demonstrated significant performance improvements over general feature adaptation across all tasks \textbf{(Figure 4d,e)}. For example, in the three special staining tasks for glomerulus, CONCH with PathFiT enabled improved by 16.42\% ($p=5.35\times 10^{-4}$), 13.25\% ($p=2.21\times10^{-4}$), and 10.93\% ($p=1.41\times 10^{-3}$), while  UNI with PathFiT enabled improved by 15.63\% ($p=6.38\times10^{-3}$), 8.82\% ($p=1.59\times10^{-3}$), and 16.15\% ($p=1.38\times10^{-4}$). Similarly, UMAP-based visualization of features on a 2D plane revealed that, with PathFiT enabled, CONCH and UNI achieved significant separation between different categories across diverse image domains \textbf{(Extended Data Figure 7)}. To explore the interpretability of foundation models on different imaging modalities, we visualized the self-attention weights in Masson, PAS, and PASM-stained images of the same glomerulus \textbf{(Figure 4f,g)}. The results showed that PathFiT allocated more attention weights to the internal structures, indicating that the extra parameters integrated into the foundation models helped enhance the focus on more relevant morphological signals. This phenomenon was also observed in tasks across other domains \textbf{(Extended Data Figure 8)}.

\noindent\textbf{\large{Discussion}}

In this work, we introduced PathFiT, a dynamic feature learning method designed to unlock the adaptability and enhance the performance of foundation models across diverse computational pathology tasks. PathFiT dynamically re-embedded image features by adding extra parameters to the foundation model and performing backpropagation jointly with the downstream predictor. It retained the original knowledge of the foundation model while preserving its structure, enabling a plug-and-play activation and deactivation mode on top of traditional general feature learning. We then collected and established a large-scale pathology image benchmark comprising 35 clinically relevant tasks to evaluate the capabilities of PathFiT. This benchmark encompassed fine-grained classification, rare disease detection, and specialized pathology imaging analysis tasks spanning 6 imaging modalities. Our quantitative experiments demonstrated that PathFiT achieved state-of-the-art performance compared to general feature learning methods across H\&E-stained ROI, H\&E-stained WSI, special staining image, and multiple optical image tasks. Moreover, through feature visualization and heatmap distributions, we revealed that this dynamic feature learning approach offered a more specific embedding space to distinguish pathological images and improved attention to lesion areas.

There are 4 points worth noting for PathFiT. First, we observed that PathFiT significantly improved performance in tasks involving special staining and multiple optical imaging. Especially CONCH, which showed over 10\% improvement in 6 out of 7 tasks. This may be due to the fact that visual-language foundation models lack image augmentation techniques during pretraining, whereas enabling PathFiT allows for dynamic adjustment of the original features, enhancing the ability to capture signals from the images themselves. In future work, we plan to incorporate robust image augmentation strategies to optimize visual-language contrastive learning. Second, we observed that PathFiT obtained greater improvements in ROI-level tasks compared to slide-level tasks. One possible reason is that ROIs, in terms of image resolutions and cropped field of view, are closer to the pertaining patches, making it easier for the foundation model with PathFiT enabled to dynamically adjust features into a more appropriate embedding space. However, the high-resolution characteristic of WSIs requires a trade-off between the intensity of dynamic re-embedding and computational overhead. Third, we highlighted that PathFiT can also be integrated into slide-level foundation models\cite{wang2024pathology,xu2024whole,shaikovski2024prism,ding2024multimodal}, which further demonstrated the versatility and effectiveness of PathFiT. For instance, enabling PathFiT in CHIEF\cite{wang2024pathology} and LongNet\cite{xu2024whole} led to improvements of 8.27\% and 14.80\% in BRCA coarse-grained subtyping \textbf{(Extended Data Figure 9)}. Finally, we observed that PathFiT demonstrated high parameter efficiency. For example, compared to full-parameter learning, PathFiT only requires adjusting an average of 3.00\% of the parameters in patch-level foundation models \textbf{(Extended Data Figure 10a-c)} and 5.84\% in slide-level foundation models \textbf{(Extended Data Figure 10d-f)}. This efficiency makes PathFiT not only computationally friendly but also capable of quickly adapting to new tasks and datasets. We are interested in exploring the potential of PathFiT in developing advanced foundation models for subspecialties (such as glioma\cite{kondepudi2024foundation}) and multimodal imaging (such as high dimensional vectorial imaging\cite{he2021polarisation,he2022towards,he2023vectorial}).

Overall, PathFiT unlocked exceptional capabilities for pathology foundation models with dynamic feature learning. With the rapid advancement of digital pathology and precision medicine, foundation models empowered by PathFiT will offer transformative potential for clinical practice. By seamlessly adapting to diverse clinical tasks and even extending to different regions or institutions, these models can set a new benchmark for performance, ultimately reshaping the future of pathology and driving the next era of AI-powered healthcare.

\noindent\textbf{\large{Methods}}

\noindent\textbf{Adding extra parameters into pathology foundation models}

PathFiT uses LoRA\cite{hu2022lora} as extra parameters of the foundation models to dynamically adjust image features. We assume that the weight updates during the adaptation process have a lower intrinsic dimension. Although the input embeddings are projected to a smaller subspace, they can still learn the intrinsic representation. Each self-attention layer in the transformer-based pathology foundation model contains four dense linear transformation layers. Considering the weight matrix $W_{0} \in \mathbb{R}^{d_2\times d_1}$ of each linear transformation layer (ignoring bias), where $d_1$ and $d_2$ represent the dimension of the input and output embedding $x$ and $h$. We add low-rank decomposition $\Delta W$ to modify the output inside the model, as shown below:
\begin{equation*}
    h = W_{0}x + \alpha \Delta W x = W_{0}x + \alpha BAx   
\end{equation*}
where $A \in \mathbb{R}^{r\times d_1}$, $B \in \mathbb{R}^{d_2\times r}$, $r$ represents the rank value, and $\alpha$ represents the scaling value. When enabling PathFiT, $W_0$ is frozen and will not be updated, while $A$ and $B$ are trainable matrices, their parameters are updated after each back-propagation. Following the original LoRA setting, we use random Gaussian initialization for $A$ and zero matrix initialization for $B$ so that $\Delta W$ is $0$ at the beginning of fine-tuning and gradient updates can be performed.

\noindent\textbf{Downstream tasks and evaluation settings}

We evaluated the capabilities and adaptability of PathFiT across 35 tasks on two representative foundation models in computational pathology: CONCH\cite{lu2024visual} and UNI\cite{chen2024towards}. These tasks include supervised H\&E-stained ROI-level classification, vision-language contrastive prompt classification, ROI segmentation, H\&E-stained WSI tasks, and specialized pathology imaging classification. To align the model structures of CONCH and UNI, we remove the vision-text alignment layer from CONCH and use its vision tower as the backbone. The $r$ and $\alpha$ parameters are fixed at $64$ and $1$ to eliminate the need for parameter tuning. We use the official pretrained weights of CONCH\footnote{\href{https://huggingface.co/MahmoodLab/CONCH}{huggingface.co/MahmoodLab/CONCH}} and UNI\footnote{\href{https://huggingface.co/MahmoodLab/UNI}{huggingface.co/MahmoodLab/UNI}}. The details of these tasks are described below.  

\noindent\textbf{Supervised ROI classification.} We compare the performance of CONCH and UNI between disabling and enabling PathFiT. We use a single linear layer (dimension of 768 for CONCH and 1024 for UNI) after foundation models to perform the classification. The batch size is set to 16. The Adam optimizer with weight decay is used, configured with a weight decay of $10^{-4}$, betas ranging from $0.9$ to $0.98$, epsilon is set to $10^{-8}$, and the learning rate is set to $10^{-4}$. Optimization is performed over 15 epochs using the cross-entropy loss function with five random seeds.

\noindent\textbf{Few-shot ROI classification with text prompt learning.} For CONCH, we connect the final layer of the foundation model to a single linear projection layer and use the text tower from the OpenAI CLIP\cite{radford2021learning} ViT-B/16 pretrained model as the text encoder. For UNI, we use the corresponding VIT-L/14 version. For each class, we convert the label into a prompt sentence: "This is a histopathological image of [CLASS]" and input it into the text encoder to obtain the corresponding text embedding. Following standard practices in machine learning\cite{zhou2022learning,zhou2022conditional}, the cosine similarity between the text embeddings and the image embeddings is computed, and the resulting probability scores are optimized using the cross-entropy loss. All other hyperparameter settings remain consistent with the settings for ROI classification.

\noindent\textbf{ROI segmentation.} Following U-Net\cite{ronneberger2015u} structure and its variants\cite{chen2024transunet}, we construct the encoder and decoder with four layers of convolution and deconvolution respectively. The image is input in parallel by the encoder and the foundation model. The image embeddings generated by the foundation model are fed into the first layer of the decoder, while the remaining layers use skip connections to combine encoder and decoder features. A hybrid loss function combining cross-entropy and dice loss (weighted equally) is used to balance pixel-wise classification accuracy and segmentation overlap quality. All other hyperparameter settings remain consistent with the settings for ROI classification.

\noindent\textbf{Weakly-supervised WSI classification.} All WSIs are processed at $20\times$ magnification, with non-overlapping tissue patches extracted using a color threshold exclusion rule. When PathFiT is enabled, all patches ($N$) in biopsy slides from PANDA and XJH are fed into the foundation model with extra parameters, generating an $N \times C$ feature matrix. This matrix is subsequently aggregated using a popular ABMIL\cite{ilse2018attention} paradigm and passes through a classification head to output class probabilities. For gigapixel resection slides, which represent the majority of cases, $64$ patches are randomly selected per iteration and fed into the foundation model with extra parameters to accommodate computational cost. The remaining patches are processed by the original foundation model, and the resulting features are concatenated and input into the ABMIL aggregator. When PathFiT is disabled, only the aggregator and classification head are updated, which is consistent with the standard two-stage MIL paradigm. We use the learning rate of $6\times10^{-4}$, 10 training epochs, and three random seeds. All other hyperparameter settings remain consistent with the settings for ROI classification.

\noindent\textbf{Details of experiment settings}

\noindent\textbf{BRCA subtyping (BACH)}\cite{aresta2019bach} is a ROI-level dataset containing 400 H\&E stained breast histology microscopy images, including four categories: normal, benign, in situ carcinoma, and invasive carcinoma. We resize images to pixels of 256 by 256, 512 by 512, 768 by 768, and 1024 by 1024, and label-stratify the train-val-test set into 0.56:0.14:0.30 for experiments.

\noindent\textbf{BRACS subtyping (BRACS)}\cite{brancati2022bracs} is a large cohort of annotated H\&E stained images to characterize breast carcinoma subtyping. It contains 547 WSIs and 4539 ROIs extracted from the WSIs, including three coarse-grained categories: benign tumors, atypical tumors, and malignant tumors, and seven fine-grained categories: normal, pathological benign, usual ductal hyperplasia, flat epithelial atypia, atypical ductal hyperplasia, ductal carcinoma in situ, and invasive carcinoma. We resize the ROI images to pixels of 256 by 256, 512 by 512, 768 by 768, and 1024 by 1024 for 7-class evaluation and perform 3-class and 7-class experiments on slide-level tasks. All experiments are conducted with the official train-val-test split.

\noindent\textbf{CRC MSI prediction (CRC-MSI)}\cite{kather2020histological} is a colorectal cancer H\&E stained ROI-level dataset from TCGA, which includes two categories: high-level MSI and non-MSI (low-level MSI and MSS). Given that the categories of the official test set are extremely unbalanced, we use the official train set, label-stratify the official train set into the train-test fold of 0.8:0.2 (15645:3912), and use the raw image size of 512 by 512 pixels for experiments.

\noindent\textbf{CRC tissue classification (CRC-100K)}\cite{kather2019predicting} is a ROI-level dataset containing 100000 human colorectal cancer and normal tissue images. It contains nine categories: adipose, background, debris, lymphocytes, mucus, smooth muscle, normal colon mucosa, cancer-associated stroma, and colorectal adenocarcinoma epithelium. We label-stratify the official NCT-CRC-HE-100K set to 0.8:0.2 as the train-val fold and use CRC-VAL-HE-7K as the test fold. All experiments are conducted using the raw image size of 224 by 224 pixels.

\noindent\textbf{Pan-cancer classification (TCGA)} is a ROI-level dataset containing 271710 H\&E stained histological images ($0.5 \mu m/pixel$) extracted from TCGA, containing 32 categories. We label-stratify it into the train-val-test fold of 0.56:0.14:0.30 (152144:38053:81513) and use the raw image size of 256 by 256 pixels for experiments.

\noindent\textbf{GI tumor tissue classification (KatherData)}\cite{kather2019deep} is a ROI-level dataset containing 11977 H\&E stained histological images for tumor detection in gastrointestinal cancer, containing 3 categories: adipose tissue and mucus (ADIMUC), stroma and muscle (STRMUS), and colorectal cancer epithelial tissue and stomach cancer epithelial tissue (TUMSTU). We label-stratify it into the train-val-test fold of 0.56:0.14:0.30 (6706:1677:3594) and use the raw image size of 512 by 512 pixels for experiments.

\noindent\textbf{GI MSI prediction (KatherMS)}\cite{kather2019deep} is a ROI-level dataset derived from gastrointestinal cancer snap-frozen samples. It contains 2 categories: microsatellite stable (MSS) and instable (MSI). We label-stratify the official train set into the train-test fold of 0.8:0.2 (48714: 12180) and use the raw image size of 224 by 224 pixels for experiments.

\noindent\textbf{OS tumor tissue classification (OTA)}\cite{arunachalam2019viable} is a ROI-level dataset composed of H\&E stained osteosarcoma histology images. It comes from the Children's Medical Center in Dallas and is collected by researchers at the University of Texas Southwestern Medical Center. The dataset consists of 1144 images with 3 categories: non-tumor, necrotic tumor, and viable tumor. We exclude images with ground truth of "viable: non-viable" and label-stratify the official train set into the train-val-test fold of 0.56:0.16:0.2 (610:153:328). All experiments are conducted using the raw image size of 1024 by 1024 pixels.

\noindent\textbf{ESCA tissue classification (TolkachData)}\cite{tolkach2023artificial} is a multi-cohort ROI-level dataset composed of H\&E stained oesophageal adenocarcinomas histology images. The dataset contains 11 categories. We use one of the cohorts (UKK1) from the University Hospital Cologne, with the train-val-test fold of 0.56:0.14:0.30 (19425:4862:10417) and use the raw image size of 256 by 256 pixels for experiments.

\noindent\textbf{Colorectal Precancer Detection (MHIST)}\cite{wei2021petri} is a ROI-level dataset composed of  H\&E stained images of colorectal polyps from the Department of Pathology and Laboratory Medicine at Dartmouth-Hitchcock Medical Center. It contains 2 categories: Hyperplastic Polyp and Sessile Serrated Adenoma. We label-stratify the official train set into the train-test fold of 0.8:0.2 (1740: 435) and use the raw image size of 224 by 224 pixels. 

\noindent\textbf{Epithelial cell segmentation (SegPath)}\cite{komura2023restaining} is a subset of the large-scale ROI-level segmentation dataset constructed by immunofluorescence restaining. It contains 26509 images and masks of epithelial cells, as a binary segmentation task of nuclei and non-cellular regions. We use the official train-val-test fold and resize the image size to 512 by 512 pixels for experiments. 

\noindent\textbf{Colon gland segmentation (Warwick-QU)}\cite{sirinukunwattana2017gland} is a ROI-level segmentation dataset containing 1585 glandular structures in 165 non-overlapping images. We use the official train-test fold and resize the image size to 224 by 224 pixels for experiments.

\noindent\textbf{Colon nuclei identification (CoNIC)}\cite{graham2024conic} is a ROI-level segmentation dataset of H\&E stained images. Each nucleus of images is assigned to one of the six categories: epithelial, lymphocyte, plasma, eosinophil, neutrophil, and connective tissue. We split the set into the train-test fold of 0.8:0.2 and use the raw image size of 256 by 256 pixels.

\noindent\textbf{OncoTree classification (TCGA)} consists of 10762 H\&E-stained FFPE diagnostic histopathology WSIs, including adrenal gland cancer, esophagogastric cancer, invasive breast cancer, ovarian cancer, thyroid cancer, bladder cancer, germ cell tumor, mature B-cell neoplasms, pancreatic cancer, uterine sarcoma, cervical cancer, glioma, melanoma, prostate cancer, colorectal cancer, head and neck cancer, mesothelioma, renal cell carcinoma, endometrial cancer, hepatobiliary cancer, non-small cell lung cancer, and thymic tumor. Based on the OncoTree cancer classification system\cite{kundra2021oncotree}, the database is further categorized into 30 OncoTree codes. We label-stratify all the data into the train-val-test fold of 0.5:0.25:0.25 (5365:2694:2703) for 30-class experiments.

\noindent\textbf{Pan-cancer classification (CPTAC)} consists of 5881 H\&E-stained FFPE diagnostic histopathology WSIs from 12 cancer types: acute myeloid leukemia, breast cancer, clear cell renal cell carcinoma, cutaneous melanoma, colon adenocarcinoma, glioblastoma multiforme, head and neck squamous cell carcinoma, lung squamous cell carcinoma, lung adenocarcinoma, ovarian cancer, pancreatic ductal adenocarcinoma, and sarcoma. We label-stratify all the data into the train-val-test fold of 0.5:0.2:0.30 (2937:1172:1772) for 12-class experiments.

\noindent\textbf{Breast metastasis fine-grained detection (Camelyon+)}\cite{ling2024towards} consists of 1350 H\&E histopathology WSIs, including 871 negative cases, 174 micro-metastasis, 251 macro-metastasis, and 54 isolated tumor cells (ITCs). These WSIs are derived from Camelyon-16\cite{bejnordi2017diagnostic} and Camelyon-17\cite{litjens20181399,bandi2018detection} grand challenge and are cleaned by professional pathologists. We label-stratify all the data into the train-val-test fold of 0.5:0.3:0.2 (675:268:407) for 4-class experiments.

\noindent\textbf{Cervical lesions detection (TissueNet)}\cite{lomenie2022can} consists of 1013 H\&E histopathology WSIs, including 268 normal or subnormal cases, 288 low-grade squamous intraepithelial lesion cases, 238 high-grade squamous intraepithelial lesion cases, and 219 invasive squamous carcinoma cases. The objective of this dataset is to detect epithelial lesions of the uterine cervix. We label-stratify all the data into the train-val-test fold of 0.5:0.2:0.3 (506:201:306) for 4-class experiments.

\noindent\textbf{Brain tumor subtyping (EBRAINS)}\cite{roetzer2022digital} consists of 2100 H\&E histopathology WSIs from the EBRAINS Digital Tumor Atlas sourced from the University of Vienna, including 47 anaplastic astrocytoma (\textit{IDH}-mutant), 47 anaplastic astrocytoma (\textit{IDH}-wildtype), 34 glioblastoma (\textit{IDH}-mutant), 469 glioblastoma (IDH-wildtype), 59 gliosarcoma, 171 pilocytic astrocytoma, 81 schwannoma, 50 anaplastic ependymoma, 96 ependymoma, 88 ganglioglioma, 59 diffuse large B-cell lymphoma of the CNS, 32 Langerhans cell histiocytosis, 46 anaplastic meningioma, 31 angiomatous meningioma, 82 atypical meningioma, 57 fibrous meningioma, 104 meningothelial meningioma, 41 secretory meningioma, 67 transitional meningioma, 87 haemangioblastoma, 30 haemangioma, 34 haemangiopericytoma, 37 lipoma, 47 metastatic tumours, 70 diffuse astrocytoma (\textit{IDH}-mutant), 85 adamantinomatous craniopharyngioma, 99 pituitary adenoma. We label-stratify all the data into the train-val-test fold of 0.5:0.2:0.3 (1044:407:649).

\noindent\textbf{Glioma \textit{IDH1} prediction and histomolecular subtyping (EBRAINS)}\cite{roetzer2022digital} consists of 692 H\&E histopathology WSIs from the EBRAINS cohort, including 123 astrocytoma, \textit{IDH1}-mutant (47 from anaplastic astrocytoma, 70 from diffuse astrocytoma, 6 from gemistocytic astrocytoma), 34 glioblastoma, \textit{IDH1}-mutant, 66 astrocytoma, \textit{IDH1}-wildtype (47 from anaplastic astrocytoma, 19 from diffuse astrocytoma), 469 glioblastoma, \textit{IDH1}-wildtype. We label-stratify all the data into the train-val-test fold of 0.5:0.2:0.3 (346:135:211) for 4-class histomolecular subtyping experiments and of 0.5:0.2:0.3 (347:137:208) for 2-class \textit{IDH1} status prediction (\textit{IDH1}-mutant vs \textit{IDH1}-wildtype) experiments.

\noindent\textbf{BRCA \textit{HER2} prediction and IHC scoring (HEROHE)}\cite{conde2021herohe} consists of 508 H\&E histopathology WSIs from the HEROHE ECDP2020 grand challenge, including 63 score 0 (negative), 65 cases of score 1 (negative), 136 cases of score 2 with positive \textit{HER2} status, 178 cases of score 2 with negative \textit{HER2} status, 66 cases of score 3 (positive). We label-stratify the official train fold with IHC scoring ground truth into the train-val fold of 0.8:0.2 (286:73), resulting in the train-val-test fold of 286:73:149 for 2-class \textit{HER2} status prediction and 4-class IHC scoring experiments.

\noindent\textbf{Pan-cancer TILs scoring (TCGA)} consists of 3727 H\&E histopathology WSIs from the TCGA cohort, including 42 cases of no obvious infiltration, 723 non-brisk multifocal cases, 640 non-brisk focal cases, 1422 brisk diffuse cases, 900 brisk band-like cases. We label-stratify the train-val-test fold of 0.5:0.2:0.3 (1863:744:1120) for 5-class TIL pattern scoring experiments.

\noindent\textbf{PRAD screening and ISUP grading (PANDA)}\cite{bulten2022artificial} consists of 5455 H\&E histopathology biopsy WSIs from Karolinska Institute and 5160 WSIs from Radboud University Medical Center, including 1924+967 G0, 1814+852 G1, 668+675 G2, 317+925 G3, 481+768 G4, 251+973 G5. They are derived from the Prostate Cancer Grade Assessment (PANDA) challenge. We label-stratify the train-val-test fold with ISUP grading ground truth of 0.5:0.2:0.3 (2726:1088:1641 and 2578:1030:1552) for the grading (G0 vs G1 vs G2 vs G3 vs G4 vs G5) and early-cancer screening (G0 vs G1+G2+G3+G4+G5) experiments. 

\noindent\textbf{cervical inflammatory tissue classification (XJH)} consists of 452 H\&E histopathology biopsy WSIs from Xijing Hospital, including 154 benign, 89 inflammation, and 209 squamous. We label-stratify the train-val fold of 0.56:0.44 (253:199) for 3-class experiments.

\noindent\textbf{Glomerular structure classification (XJH)} consists of 2069 transmission electron microscopy images extracted from 400 renal biopsy cases in Xijing Hospital. They are fixed with glutaraldehyde and osmium tetroxide, stained with uranyl acetate and lead citrate, and imaged using a Hitachi-7800 transmission electron microscope. The database is utilized to classify 19 diagnostic structural types, including 1) 109 GBM stratification, 101 thinning, 108 thickening, and 104 normal in basement membrane lesions; 2) 114 subendothelial space widening, 103 subendothelial, 104 minimal subepithelial, 112 subepithelial, and 90 subepithelial resorptions in deposits; 3) 125 mesangial deposits and 101 normal mesangial regions in mesangial area lesions; 4) 111 minor fusion, 103 partial fusion, 110 extensive fusion in foot process lesions; 5) 118 structural changes of glomeruli, 116 platelets, and 106 neutrophil aggregates in structural differentiation; 6) 131 amyloidosis nephropathy and 103 fabry nephropathy in other structural lesions. We resize the raw image size from 3296 by 2563 pixels to 1024 by 1024 pixels and label-stratify the train-val-test fold of 0.56:0.14:0.30 (1143:297:629) for 19-class experiments.

\noindent\textbf{Masson's Trichrome glomerular classification (XJH)} consists of 482 Masson-stained glomerular images extracted from histopathology biopsy WSIs in Xijing Hospital. We use pretrained Mask-R-CNN\cite{he2017mask} with swin transformer\cite{liu2021swin} to automatically segment and obtain glomeruli. We divide the stage of Mesangial hypercellularity into four classes: 200 normal, 57 early stage, 112 intermediate stage, and 113 late stage. By resizing the raw image size of 512 by 512 pixels, we label-stratify the train-val-test fold of 0.5:0.2:0.3 (268:68:146) for 4-class experiments.

\noindent\textbf{Periodic Acid-Schiff glomerular classification (XJH)} consists of 3187 PAS-stained glomerular images extracted from histopathology biopsy WSIs in Xijing Hospital. We use pretrained Mask-R-CNN\cite{he2017mask} with swin transformer\cite{liu2021swin} to automatically segment and obtain glomeruli. We divide the stages of Mesangial hypercellularity into four classes: 1200 normal, 1129 early stage, 479 intermediate stage, and 379 late stage. By resizing the raw image size of 512 by 512 pixels, we label-stratify the train-val-test fold of 0.5:0.2:0.3 (1784:446:957) for 4-class experiments.

\noindent\textbf{Periodic Acid-Schiff Methenamine glomerular classification (XJH)} consists of 498 PASM-stained glomerular images extracted from histopathology biopsy WSIs in Xijing Hospital. We use pretrained Mask-R-CNN\cite{he2017mask} with swin transformer\cite{liu2021swin} to automatically segment and obtain glomeruli. We divide the stages of Mesangial hypercellularity into four classes: 200 normal, 76 early stage, 135 intermediate stage, and 87 late stage. By resizing the raw image size of 512 by 512 pixels, we label-stratify the train-val-test fold of 0.5:0.2:0.3 (277:70:151) for 4-class experiments.

\noindent\textbf{Immunofluorescence sediment organization classification (XJH)} consists of 1711 Olympus fluorescence microscope glomerular images collected from Xijing Hospital, including 1053 capillary walls and 658 mesangial areas. The images are captured with $10\times$ magnification. We label-stratify the train-val-test fold of 0.56:0.14:0.30 (957:240:514) and resize the raw image size of 1024 by 1024 pixels for 2-class experiments.  

\noindent\textbf{Immunofluorescence deposit distribution detection (XJH)} consists of 1709 Olympus fluorescence microscope glomerular images collected from Xijing Hospital, including 747 segmental and 962 diffuse distribution. The images are captured with $10\times$ magnification. We label-stratify the train-val-test fold of 0.56:0.14:0.30 (955:240:514) and resize the raw image size of 1024 by 1024 pixels for 2-class experiments.

\noindent\textbf{Immunohistochemistry tissue classification (MIHIC)}\cite{wang2024mihic} is a patch-level dataset that consists of 309698 images across 12 different IHC stains, where six tissue types are annotated. We use the official train-val-test fold with the raw image size of 128 by 128 pixels for 6-class experiments.

\noindent\textbf{\large{Computing software and hardware}}

We conduct all experiments and analyses using Python (v3.12.2). Fine-tuning for all downstream tasks is performed on a single NVIDIA A100 GPU. All methods are implemented using the popular open-source deep learning framework PyTorch (v2.4.1, CUDA 12.1). For foundation models, we use the open-source Timm library (v1.0.9) for model definitions. We extend the official LoRA code\footnote{\href{https://github.com/microsoft/LoRA}{github.com/microsoft/LoRA}} to adapt it for vision transformers in the Timm library. For the text encoder, we use open\-clip library\footnote{\href{https://github.com/mlfoundations/open_clip}{github.com/mlfoundations/open\_clip}} and load model weights from Hugging Face\footnote{\href{https://huggingface.co/}{huggingface.co}}. For segmentation tasks, we make modifications based on TransUNet codebase\footnote{\href{https://github.com/Beckschen/TransUNet}{github.com/Beckschen/TransUNet}}. ABMIL and heatmap visualizations for WSIs are implemented using the CLAM codebase\footnote{\href{https://github.com/mahmoodlab/CLAM}{github.com/mahmoodlab/CLAM}}. WSI processing is performed with Opensdpc codebase\footnote{\href{https://github.com/WonderLandxD/opensdpc}{github.com/WonderLandxD/opensdpc}}. ROI image visualization is executed using the HIPT codebase\footnote{\href{https://github.com/mahmoodlab/HIPT}{github.com/mahmoodlab/HIPT}}. Detailed Python library versions include: matplotlib (v3.9.2), numpy (v1.26.4), open\_clip\_torch (v2.27.1), opencv-python (v4.10.0.84), opensdpc (v1.0.0), openslide-python (v1.3.1), pandas (v2.2.2), pillow (v10.4.0), scikit-learn (v1.5.2), and tqdm (v4.66.4).

\noindent\textbf{\large{Data availability}}

All publicly available datasets analyzed in this study can be accessed through their respective data portals: \href{https://iciar2018-challenge.grand-challenge.org/Dataset/}{BACH}, \href{https://www.bracs.icar.cnr.it/}{BRACS}, \href{https://zenodo.org/records/3832231}{CRC-MSI}, \href{https://zenodo.org/records/1214456}{CRC-100K}, \href{https://zenodo.org/records/5889558}{Pan-Cancer Classification}, \href{https://doi.org/10.5281/zenodo.2530789}{katherData}, \href{https://doi.org/10.5281/zenodo.2532612}{KatherMS}, \href{https://doi.org/10.7937/tcia.2019.bvhjhdas}{OTA}, \href{https://doi.org/10.5281/zenodo.7548828}{TolkachData}, \href{https://bmirds.github.io/MHIST/}{MHIST}, \href{https://dakomura.github.io/SegPath/}{SegPath}, \href{https://doi.org/10.1109/TMI.2015.2433900}{Warwick-QU}, \href{https://conic-challenge.grand-challenge.org/}{CoNIC}, \href{https://portal.gdc.cancer.gov/}{TCGA}, \href{https://proteomic.datacommons.cancer.gov/pdc/}{CPTAC}, \href{https://doi.org/10.57760/sciencedb.16442}{Camelyon+}, \href{https://www.drivendata.org/competitions/67/competition-cervical-biopsy/page/256/}{TissueNet}, \href{https://www.doi.org/10.25493/WQ48-ZGX}{EBRAINS}, \href{https://ecdp2020.grand-challenge.org/}{HEROHE}, \href{https://doi.org/10.7937/K9/TCIA.2018.Y75F9W1}{TCGA-TILs}, \href{https://panda.grand-challenge.org/}{PANDA}, \href{https://zenodo.org/records/10065510}{MIHIC}. Following institution policies, all requests for data collected or curated in-house will be evaluated case-by-case to determine whether the data requested and the use case comply with intellectual property or patient privacy obligations.

\noindent\textbf{\large{Code availability}}

Code for performing various downstream tasks using PathFiT adaptation will be released upon publication. We document all technical methods and software libraries used in the study while ensuring the paper is accessible to the broader clinical and scientific audience.

\noindent\textbf{\large{Author contributions}}

J.L., T.G., X.-W.B., Z.W., L.G., C.H., and Y.H. conceived the study and designed the experiments. J.L., Y.W., X.L., J.T., and X.W. perform model development for downstream tasks. J.L., Q.X., Jing Li, Q.H., Z.W., Z.S., Z.L., and T.C. collected the data and organized the datasets for downstream tasks. J.L., Y.W., X.L., Y.M., and Z.Z. organized the codebases for downstream tasks. J.L., T.G., and X.L. performed experimental analysis regarding H\&E-stained tasks. J.L., T.G., Y.W., Jing Li, Y.M., and Z.Z. performed experimental analysis regarding specialized imaging tasks. J.L., T.G., Q.X., Jing Li, Z.L., T.C., X.-W.B., Z.W., L.G., C.H., and Y.H. interpreted experimental results and provided feedback on the study. J.L., T.G., C.H., and Y.H. prepared the manuscript with input from all co-authors. X.-W.B., Z.W., L.G., C.H., and Y.H. supervised the research.

\noindent\textbf{\large{Acknowledgements}}

This work was supported by the National Natural Science Foundation of China (NSFC) under Grant No.82430062, the Shenzhen Engineering Research Centre (XMHT20230115004), the Shenzhen Science and Technology Innovation Commission (KCXFZ20201221173207022), and Cross-disciplinary Research and Innovation Fund Research Plan of Tsinghua Shenzhen International Graduate School under Grant No.JC2024002. Z.L. and L.G. were also supported by the Natural Science Foundation of Jiangsu Province (BK20241793) and the Suzhou Science and Technology Development Program (SKY2023009). C.H. was also supported by the St John's College, the University of Oxford, and the Royal Society (URF\textbackslash R1\textbackslash 241734). We thank the Jilin FuyuanGuan Food Group Co., Ltd for their collaboration.


\newpage

\clearpage
\begin{figure*}
\centering
\includegraphics[width=1\textwidth]{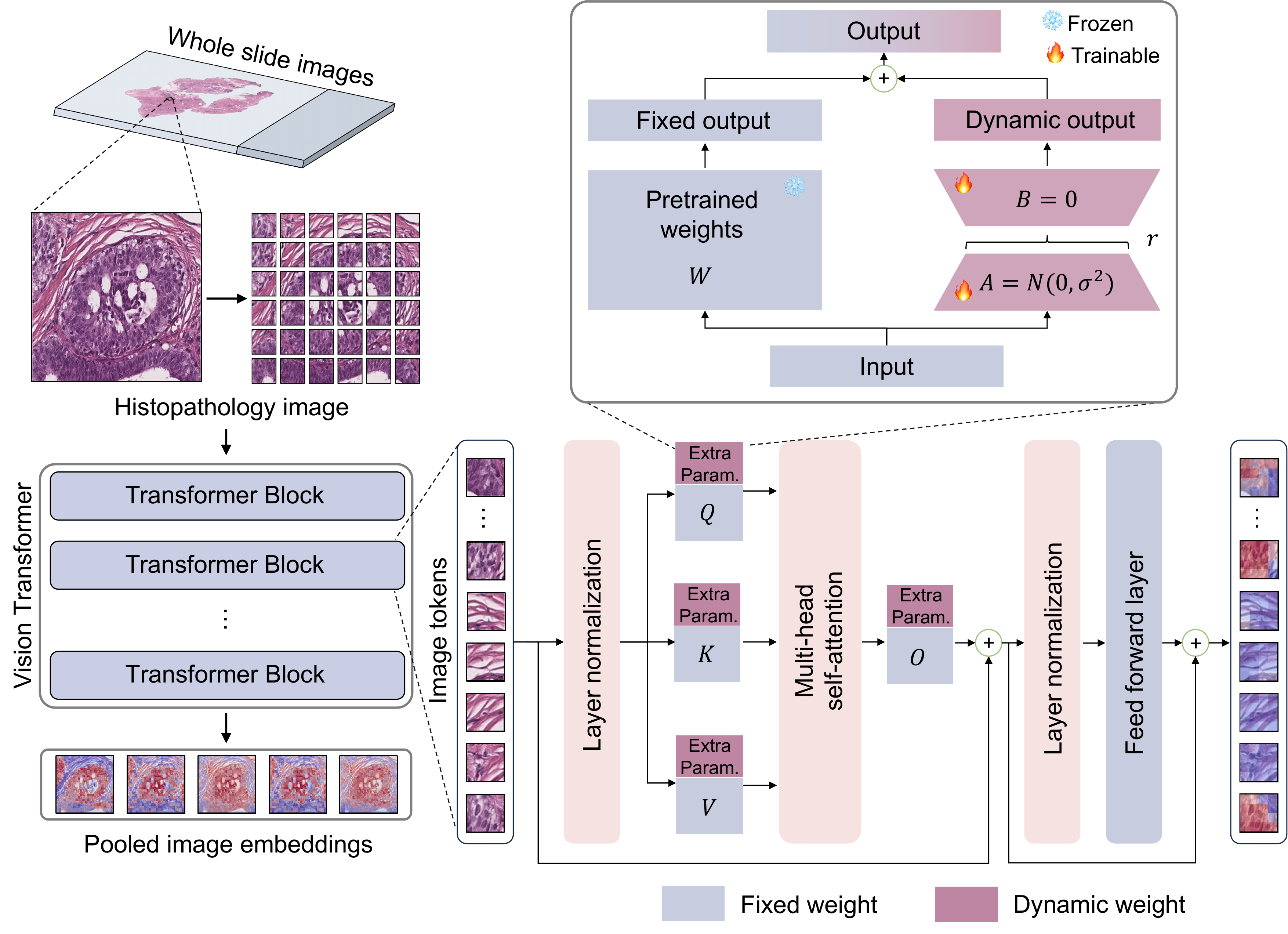}
\caption*{\textbf{Extended Data Figure 1: The architecture of plug-and-play PathFiT.} The foundation model converts pathology images into token-based sequences and inputs them into vision transformers (ViTs). PathFiT introduces extra parameters into each Transformer block of the ViT. Specifically, for the query, key, value, and output linear layers in the multi-head self-attention mechanism, matrices $A$ and $B$ are added to their paths in parallel. By jointly updating $A$ and $B$ inside the foundation model along with the task-specific classifier, PathFiT dynamically adapts general features of images to task-specific feature spaces, thereby improving model performance.}
\end{figure*}

\clearpage
\begin{figure*}
\centering
\includegraphics[width=1\textwidth]{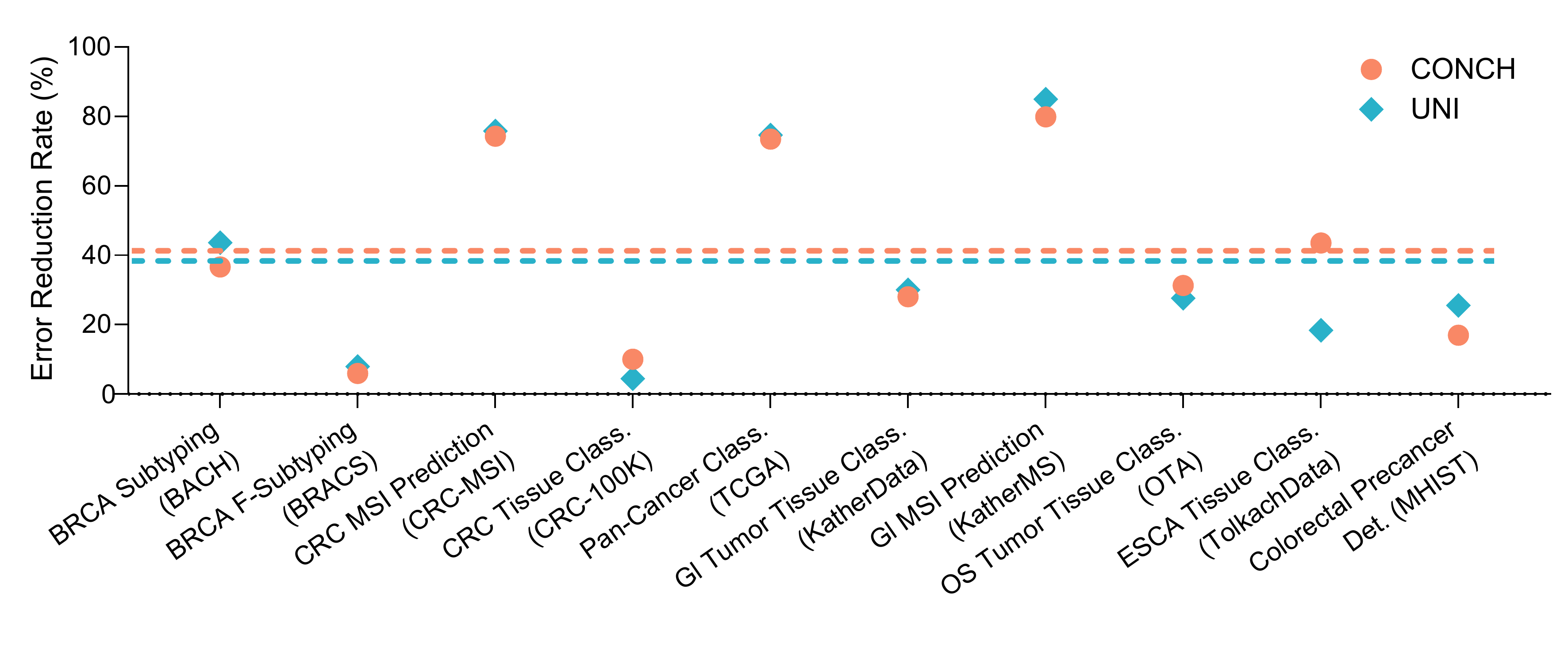}
\caption*{\textbf{Extended Data Figure 2: Results of error reduction rate across all ROI-level classification tasks by enabling PathFiT for CONCH and UNI.}}
\end{figure*}

\clearpage
\begin{figure*}
\centering
\includegraphics[width=1\textwidth]{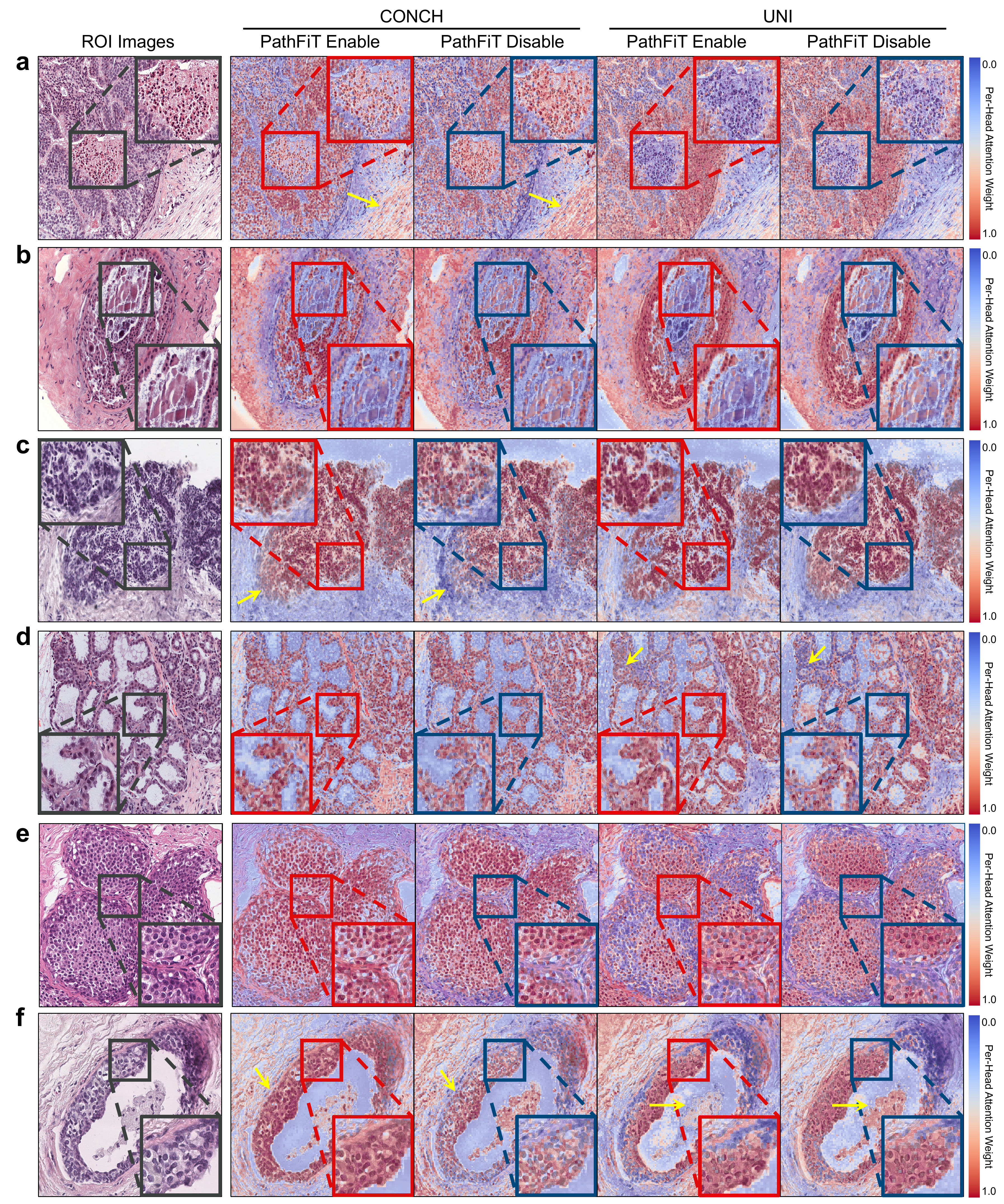}
\caption*{\textbf{Extended Data Figure 3: Visualization of multi-head self-attention in BACH and BRACS cohort.} We resize the images to a resolution of 1792 by 1792 pixels and generate heatmaps by visualizing the weight scores of the class token relative to each patch token in the final transformer layer of the foundation model, and mapping these scores to their corresponding positions in the image.}
\end{figure*}

\clearpage
\begin{figure*}
\centering
\includegraphics[width=1\textwidth]{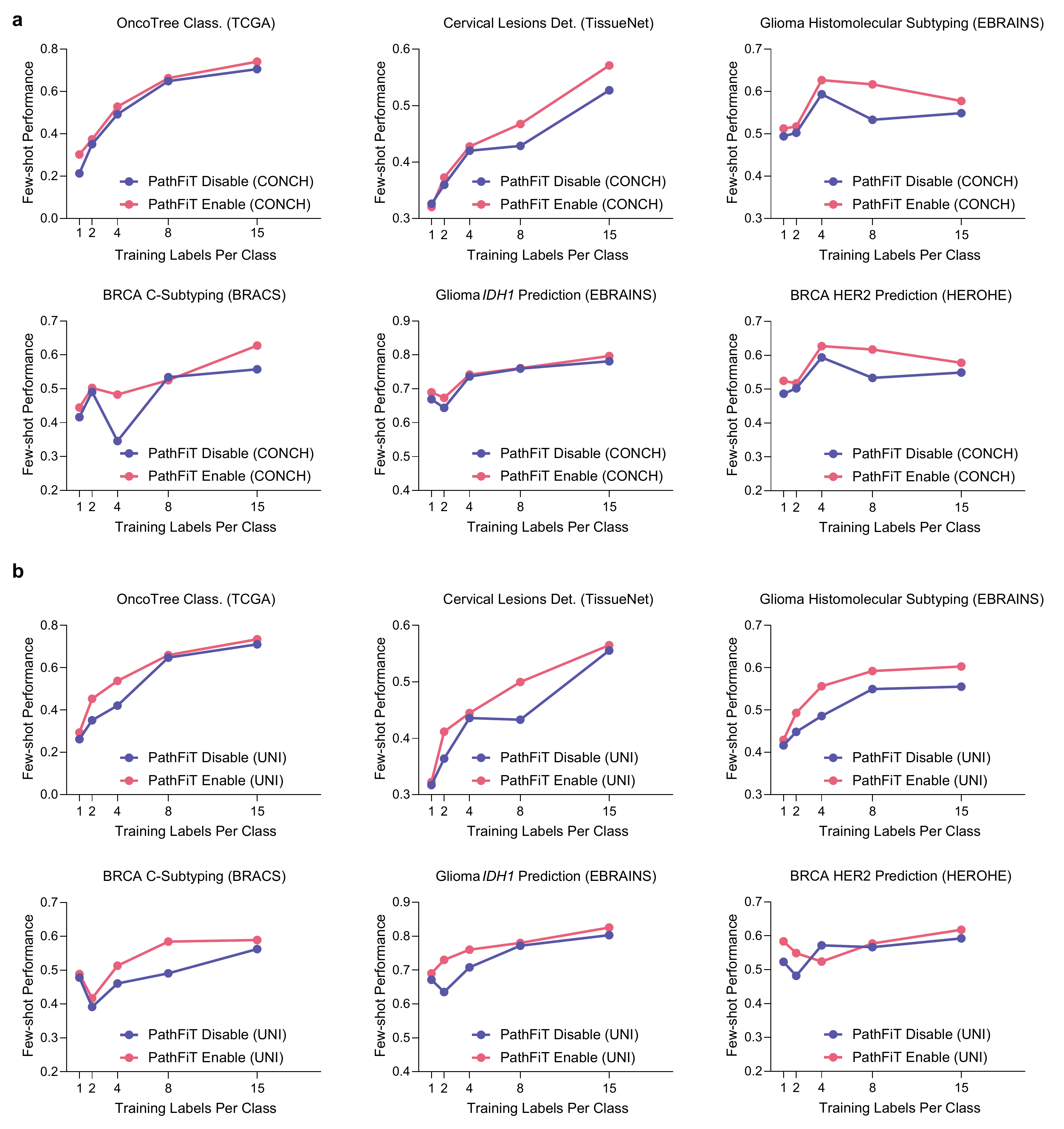}
\caption*{\textbf{Extended Data Figure 4: Few-shot slide-level classification.} We compare the performance of ABMIL fine-tuning with PathFiT to that of vanilla ABMIL fine-tuning (PathFiT Disable) in six tasks, including OncoTree classification (TCGA), cervical lesions detection (TissueNet), glioma histomolecular subtyping (EBRAINS), BRCA coarse-grained subtyping (BRACS), glioma \textit{IDH1} prediction (EBRAINS), and BRCA \textit{HER2} prediction (HEROHE) on two foundation models: \textbf{a.} CONCH and \textbf{b.} UNI.}
\end{figure*}

\clearpage
\begin{figure*}
\centering
\includegraphics[width=1\textwidth]{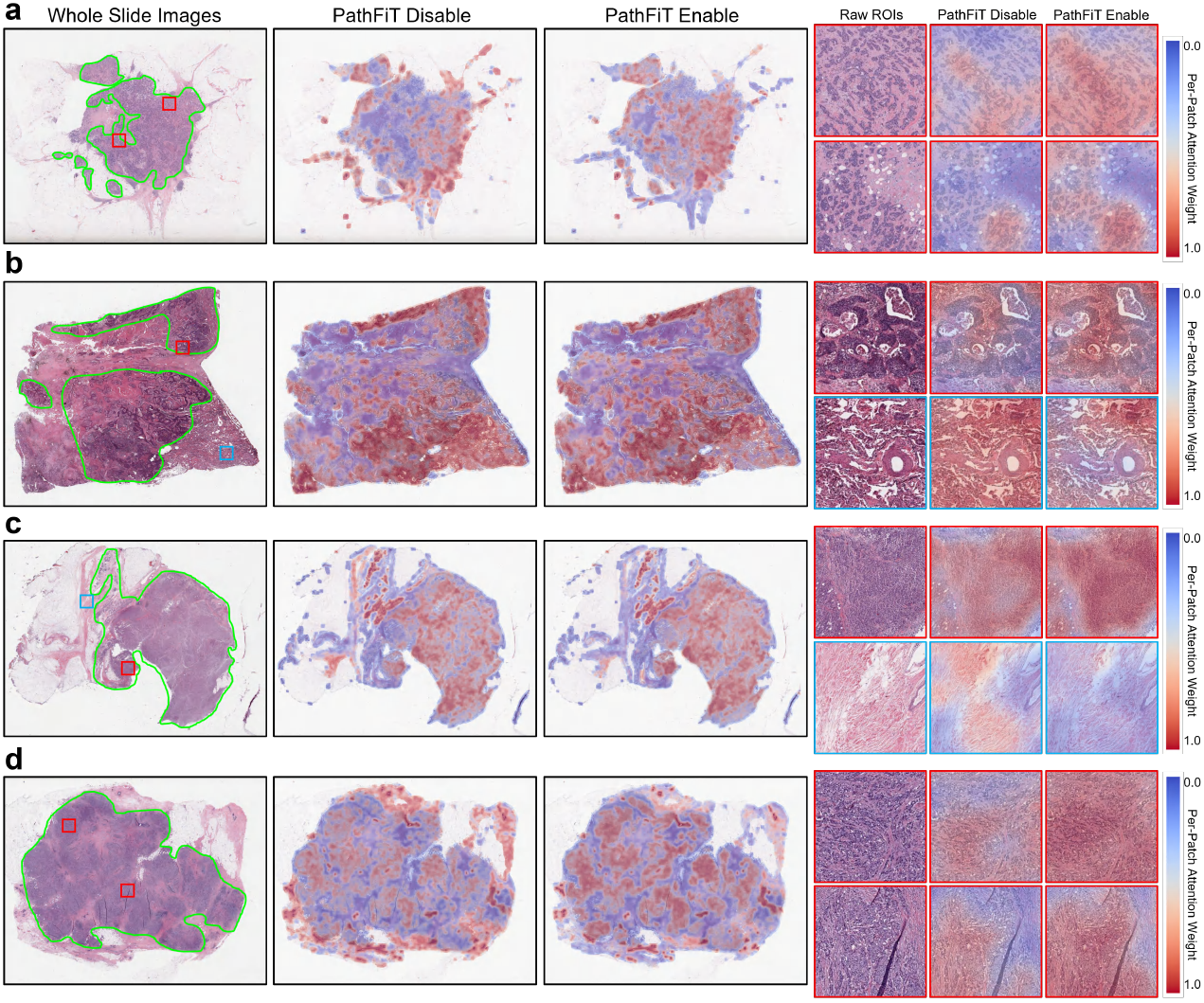}
\caption*{\textbf{Extended Data Figure 5: Visualization on WSIs with supervised slide-level classification.} We crop tissue-contain patches with 85\% overlap and map the attention weights of the ABMIL aggregator to their corresponding spatial locations, following the official CLAM visualization codebase\cite{lu2021data}. Visualization on \textbf{a.} BRCA, \textbf{b.} LUSC, \textbf{c.} BRCA, \textbf{d.} BRCA from the TCGA cohort (green outlines indicate the annotated cancerous regions) reveals that the aggregator with PathFiT enabled focuses on a broader range of cancerous regions than with PathFiT disabled. PathFiT also optimizes the local weight probability distribution, evident in selected ROIs (blue boxes indicate non-cancerous regions and red boxes indicate regions within cancerous areas).}
\end{figure*}

\clearpage
\begin{figure*}
\centering
\includegraphics[width=1\textwidth]{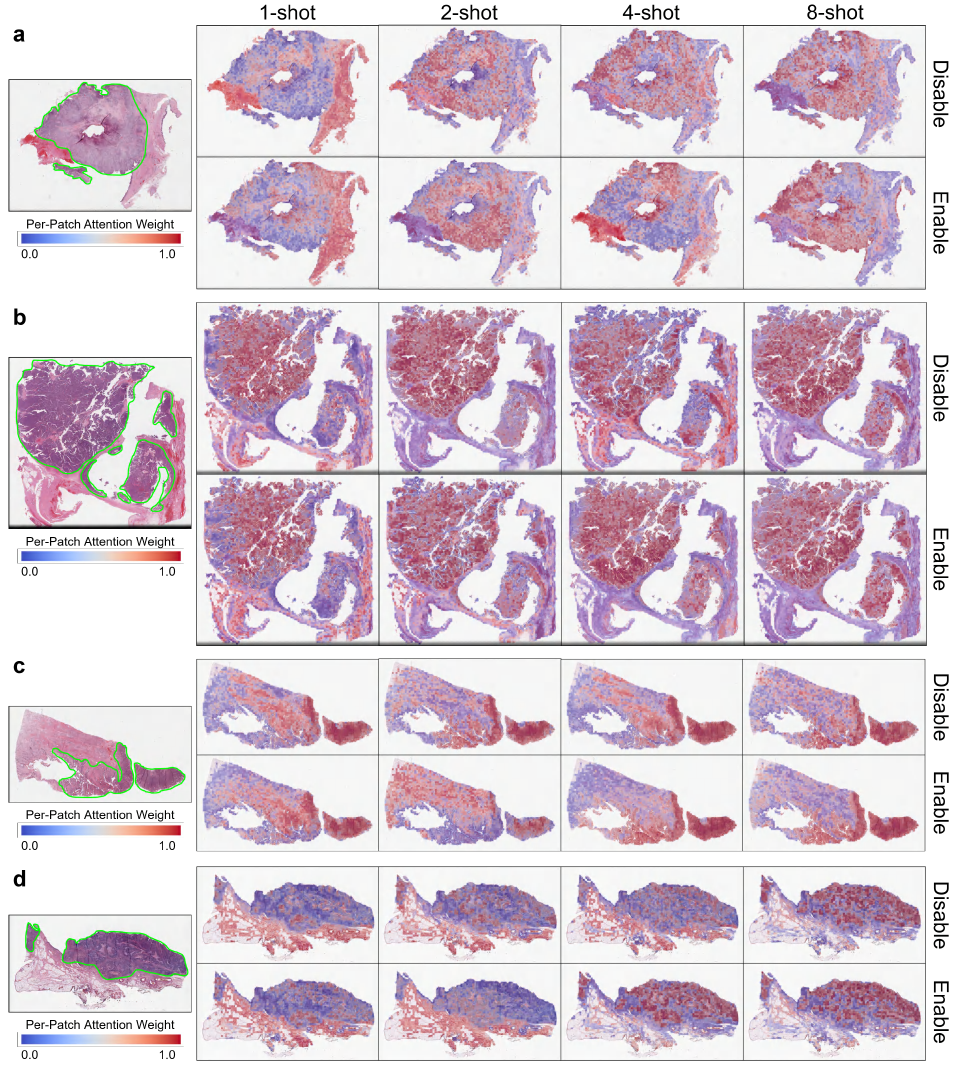}
\caption*{\textbf{Extended Data Figure 6: Visualization on WSIs with few-shot slide-level classification.} By using the official CLAM visualization codebase\cite{lu2021data}, we use different shot settings to visualize the heatmap of \textbf{a.} KIRC, \textbf{b.} OV, \textbf{c.} PAAD, and \textbf{d.} STAD from the TCGA cohort using CONCH-based ABMIL. We observe that as the number of shots increases, high attention scores increasingly focus on cancerous regions. Furthermore, weights with PathFiT enabled demonstrate a well-distributed attention pattern with fewer shots than with PathFiT disabled.}
\end{figure*}

\clearpage
\begin{figure*}
\centering
\includegraphics[width=1\textwidth]{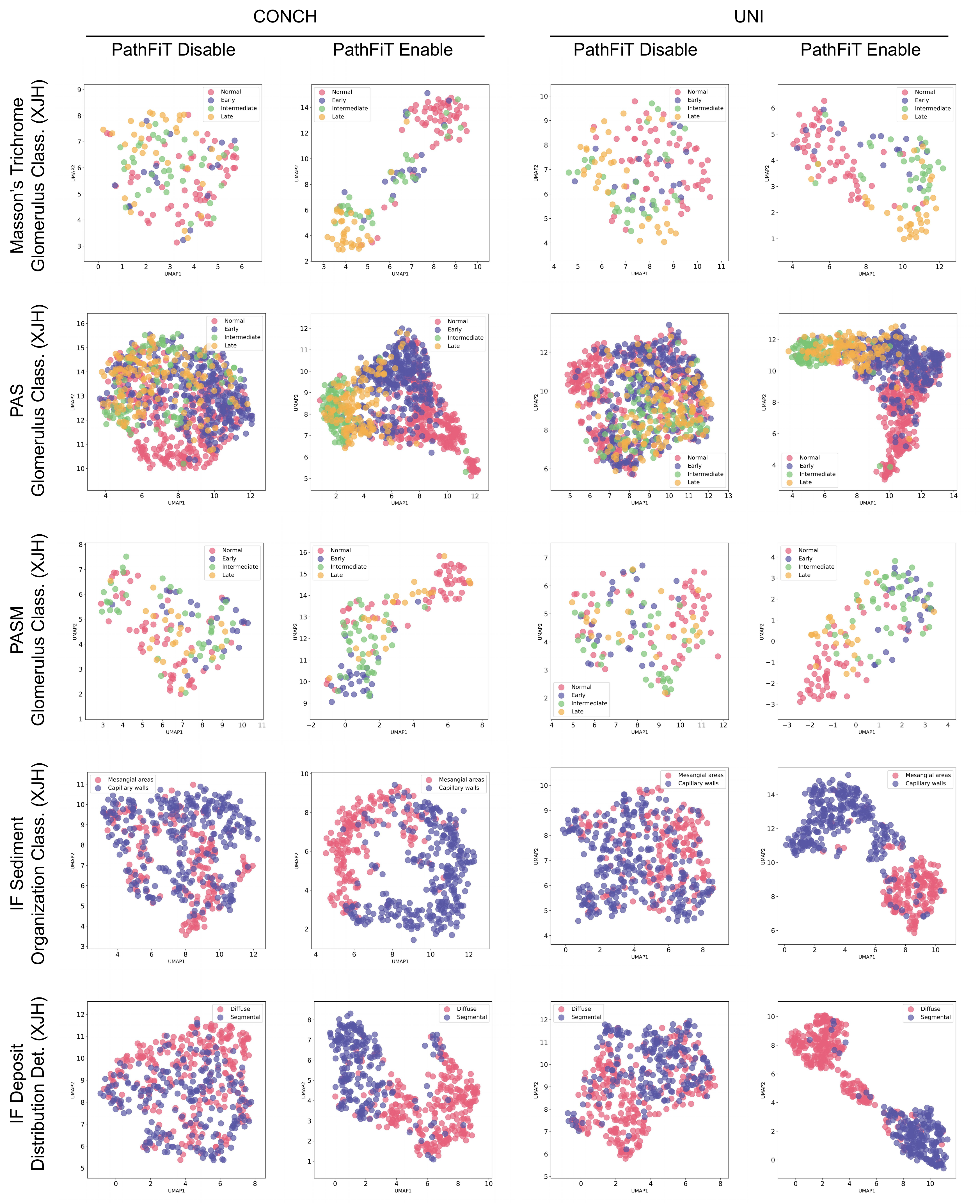}
\caption*{\textbf{Extended Data Figure 7: Comparison results of 2D visualization of image embeddings.} We visualize the UMAP comparison generated by image embedding between disabling and enabling PathFiT. Across five tasks involving Masson, PAS, PASM, and immunofluorescence staining, we observe that the feature embeddings achieve greater separation between different classes with PathFiT enabled.}
\end{figure*}

\clearpage
\begin{figure*}
\centering
\includegraphics[width=1\textwidth]{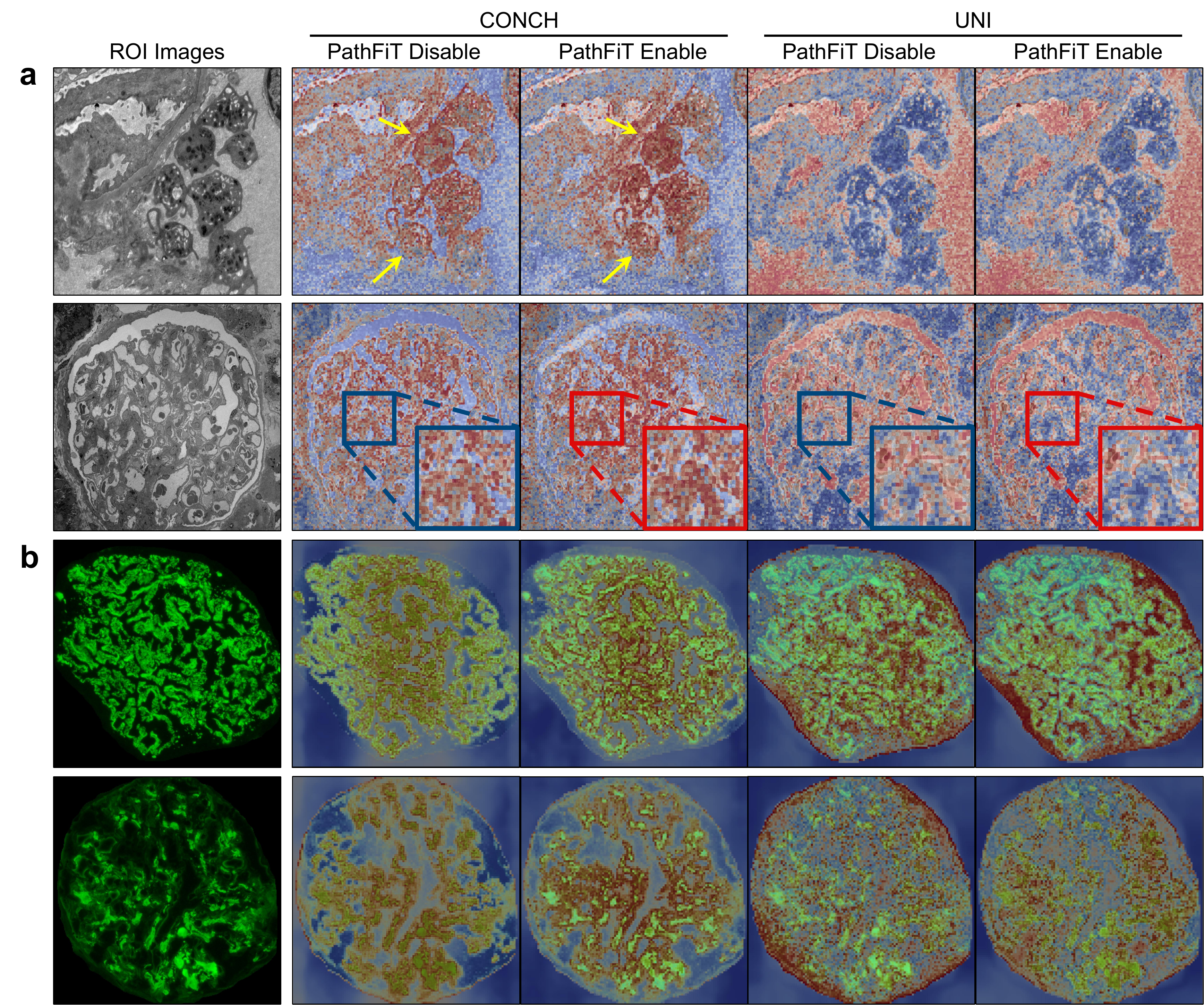}
\caption*{\textbf{Extended Data Figure 8: Visualization of multi-head self-attention in specialized pathology imaging tasks.} We resize the \textbf{a.} transmission electron microscopy and \textbf{b.} immunofluorescence images to a resolution of 1792 by 1792 pixels and generate heatmaps by visualizing the weight scores of the class token in the self-attention. We demonstrate that PathFiT adaptation allocates more attention weights to intraglomerular structures.}
\end{figure*}

\clearpage
\begin{figure*}
\centering
\includegraphics[width=1\textwidth]{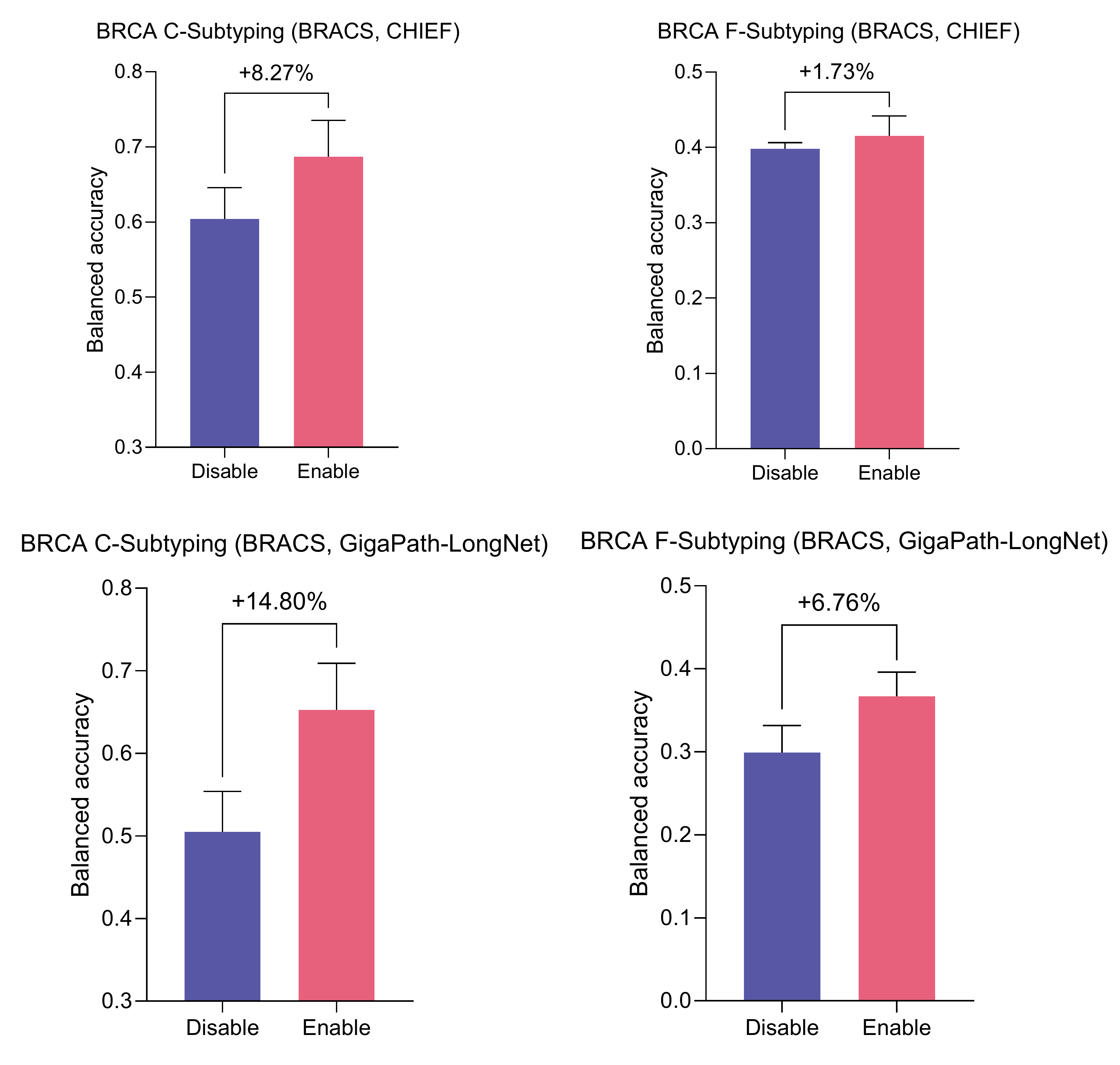}
\caption*{\textbf{Extended Data Figure 9: Comparison results on BRCA fine-grained and coarse-grained subtyping with slide-level foundation model CHIEF\cite{wang2024pathology} and GigaPath-LongNet\cite{xu2024whole} between disabling and enabling PathFiT.}}
\end{figure*}

\clearpage
\begin{figure*}
\centering
\includegraphics[width=1\textwidth]{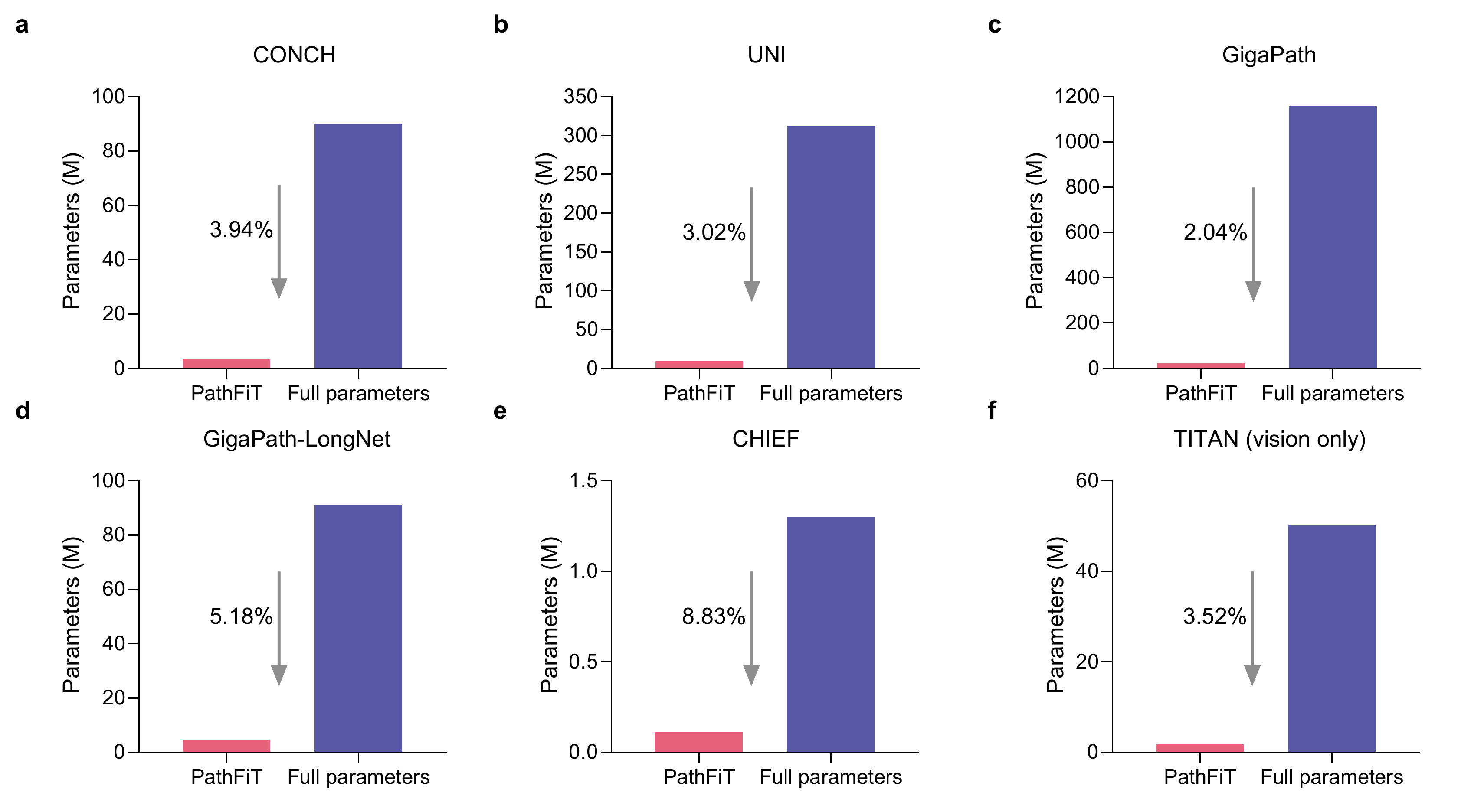}
\caption*{\textbf{Extended Data Figure 10: Comparison of the parameters of PathFiT enabled with the full parameters in six foundation models of computational pathology: CONCH\cite{lu2024visual}, UNI\cite{chen2024towards}, GigaPath\cite{xu2024whole}, GigaPath-LongNet\cite{xu2024whole}, CHIEF\cite{wang2024pathology}, and TITAN\cite{ding2024multimodal}.}}
\end{figure*}

\clearpage
\begin{table}
\centering
\begin{tabular}{lllll}
\toprule
Model &  Adaptation &  Balanced accuracy &  ROC AUC    & Weighted F1\\
\midrule
CONCH & PathFiT Disable & 78.50 (75.65-81.35) & 94.84 (93.79-95.89) & 78.07 (75.15-80.98) \\
      & PathFiT Enable & \textbf{86.50 (85.09-87.91)} & \textbf{97.70 (97.18-98.22)} & \textbf{86.43 (84.98-87.87)} \\
\midrule
UNI   & PathFiT Disable & 85.67 (84.08-87.25)	& 96.63 (96.01-97.26) & 85.45 (83.85-87.05) \\
      & PathFiT Enable & \textbf{92.17 (90.94-93.39)}	& \textbf{99.10 (98.93-99.28)} & \textbf{92.14 (90.97-93.32)}\\
\bottomrule
\end{tabular}
\caption{\textbf{ROI-level supervised classification results on BRCA subtyping (BACH)} in terms of balanced accuracy, ROC AUC, and weighted F1 score. The best-performing model for each metric is bolded, with 95\% CI in parentheses.}
\label{tab:bach_full}
\end{table}

\begin{table}
\centering
\begin{tabular}{lllll}
\toprule
Model &  Adaptation &  Balanced accuracy &  ROC AUC    & Weighted F1\\
\midrule
CONCH & PathFiT Disable & 62.21 (61.96-62.45)	& 90.80 (90.64-90.96) & 60.99 (60.66-61.33) \\
      & PathFiT Enable & \textbf{64.47 (63.67-65.26)} & \textbf{91.89 (91.44-92.34)} & \textbf{64.29 (63.52-65.06)} \\
\midrule
UNI   & PathFiT Disable & 62.29 (61.04-63.55) & 91.25 (90.94-91.55) & 62.04 (60.69-63.39)\\
      & PathFiT Enable & \textbf{65.36 (64.30-66.42)} & \textbf{92.61 (92.38-92.83)} & \textbf{65.00 (64.00-65.99)}\\
\bottomrule
\end{tabular}
\caption{\textbf{ROI-level supervised classification results on BRCA fine-grained subtyping (BRACS)} in terms of balanced accuracy, ROC AUC, and weighted F1 score. The best-performing model for each metric is bolded, with 95\% CI in parentheses.}
\label{tab:bracs_full}
\end{table}

\begin{table}
\centering
\begin{tabular}{lllll}
\toprule
Model &  Adaptation &  Balanced accuracy &  ROC AUC    & Weighted F1\\
\midrule
CONCH & PathFiT Disable & 86.24 (86.02-86.46) & 93.66 (93.57-93.76) & 86.24 (86.02-86.46) \\
      & PathFiT Enable & \textbf{96.47 (96.16-96.79)} & \textbf{99.46 (99.41-99.52)} & \textbf{96.47 (96.16-96.79)} \\
\midrule
UNI   & PathFiT Disable & 88.74 (88.58-88.89) & 95.72 (95.67-95.77) & 88.74 (88.58-88.89)\\
      & PathFiT Enable & \textbf{97.27 (97.18-97.36)} & \textbf{99.64 (99.60-99.68)} & \textbf{97.27 (97.18-97.36)}\\
\bottomrule
\end{tabular}
\caption{\textbf{ROI-level supervised classification results on CRC MSI prediction (CRC-MSI)} in terms of balanced accuracy, ROC AUC, and weighted F1 score. The best-performing model for each metric is bolded, with 95\% CI in parentheses.}
\label{tab:crc_msi_full}
\end{table}

\begin{table}
\centering
\begin{tabular}{lllll}
\toprule
Model &  Adaptation &  Balanced accuracy &  ROC AUC    & Weighted F1\\
\midrule
CONCH & PathFiT Disable & 94.43 (94.32-94.55) & \textbf{99.66 (99.64-99.67) }& 96.08 (96.01-96.14)
 \\
      & PathFiT Enable & \textbf{94.99 (94.71-95.28)} & 99.65 (99.57-99.73) & \textbf{96.47 (96.13-96.82)}
 \\
\midrule
UNI   & PathFiT Disable & 94.07 (93.8-94.33) & \textbf{99.50 (99.48-99.52) }& 95.40 (95.11-95.69)
\\
      & PathFiT Enable & \textbf{94.33 (93.66-94.99)} & 99.38 (99.29-99.47) & \textbf{95.66 (95.12-96.21)}\\
\bottomrule
\end{tabular}
\caption{\textbf{ROI-level supervised classification results on CRC tissue (CRC-100K)} in terms of balanced accuracy, ROC AUC, and weighted F1 score. The best-performing model for each metric is bolded, with 95\% CI in parentheses.}
\label{tab:crc_100k_full}
\end{table}

\begin{table}
\centering
\begin{tabular}{lllll}
\toprule
Model &  Adaptation &  Balanced accuracy &  ROC AUC    & Weighted F1\\
\midrule
CONCH & PathFiT Disable & 81.36 (81.14-81.58) & 99.34 (99.33-99.36) & 84.49 (84.33-84.65)
 \\
      & PathFiT Enable & \textbf{95.06 (94.8-95.31)} & \textbf{99.94 (99.94-99.95) }& \textbf{95.73 (95.49-95.97)} \\
\midrule
UNI   & PathFiT Disable & 87.19 (87.04-87.34) & 99.68 (99.68-99.69) & 89.25 (89.16-89.35) \\
      & PathFiT Enable & \textbf{96.74 (96.2-97.29)} & \textbf{99.97 (99.97-99.98)} & \textbf{97.26 (96.94-97.57)}\\
\bottomrule
\end{tabular}
\caption{\textbf{ROI-level supervised classification results on pan-cancer (TCGA)} in terms of balanced accuracy, ROC AUC, and weighted F1 score. The best-performing model for each metric is bolded, with 95\% CI in parentheses.}
\label{tab:tcga_uniform_full}
\end{table}

\begin{table}
\centering
\begin{tabular}{lllll}
\toprule
Model &  Adaptation &  Balanced accuracy &  ROC AUC    & Weighted F1\\
\midrule
CONCH & PathFiT Disable & 99.86 (99.86-99.86) & \textbf{100.00 (100.00-100.00)} & 99.86 (99.86-99.86) \\
      & PathFiT Enable & \textbf{99.90 (99.86-99.94)} & \textbf{100.00 (100.00-100.00)} & \textbf{99.90 (99.86-99.94)} \\
\midrule
UNI   & PathFiT Disable & 99.89 (99.89-99.89) & \textbf{100.00 (100.00-100.00)} & 99.89 (99.89-99.89) \\
      & PathFiT Enable & \textbf{99.92 (99.91-99.93)} & \textbf{100.00 (100.00-100.00)} & \textbf{99.92 (99.91-99.93)}\\
\bottomrule
\end{tabular}
\caption{\textbf{ROI-level supervised classification results on glioma tumor tissue (KatherData)} in terms of balanced accuracy, ROC AUC, and weighted F1 score. The best-performing model for each metric is bolded, with 95\% CI in parentheses.}
\label{tab:katherdata_full}
\end{table}

\begin{table}
\centering
\begin{tabular}{lllll}
\toprule
Model &  Adaptation &  Balanced accuracy &  ROC AUC    & Weighted F1\\
\midrule
CONCH & PathFiT Disable & 84.87 (84.78-84.95) & 92.51 (92.48-92.53) & 84.86 (84.78-84.95) \\
      & PathFiT Enable & \textbf{96.95 (96.82-97.08)} & \textbf{99.59 (99.56-99.63) }& \textbf{96.95 (96.82-97.08)}\\
\midrule
UNI   & PathFiT Disable & 88.39 (88.29-88.5) & 95.13 (95.07-95.18) & 88.39 (88.28-88.50) \\
      & PathFiT Enable & \textbf{98.26 (98.19-98.32)} & \textbf{99.85 (99.84-99.86) }& \textbf{98.26 (98.19-98.32)}\\
\bottomrule
\end{tabular}
\caption{\textbf{ROI-level supervised classification results on glioma MSI status prediction (KatherMS)} in terms of balanced accuracy, ROC AUC, and weighted F1 score. The best-performing model for each metric is bolded, with 95\% CI in parentheses.}
\label{tab:kathermsi_full}
\end{table}

\begin{table}
\centering
\begin{tabular}{lllll}
\toprule
Model &  Adaptation &  Balanced accuracy &  ROC AUC    & Weighted F1\\
\midrule
CONCH & PathFiT Disable & 93.15 (92.28-94.02) & 99.10 (98.85-99.36) & 94.17 (93.39-94.95) \\
      & PathFiT Enable & \textbf{95.37 (94.88-95.85)} & \textbf{99.48 (99.43-99.52)} & \textbf{95.67 (95.17-96.18)}\\
\midrule
UNI   & PathFiT Disable & 93.40 (92.83-93.97) & 99.19 (99.08-99.31) & 94.72 (94.23-95.21)\\
      & PathFiT Enable & \textbf{95.28 (94.54-96.01)} & \textbf{99.47 (99.42-99.52)} & \textbf{95.55 (94.83-96.27)}\\
\bottomrule
\end{tabular}
\caption{\textbf{ROI-level supervised classification results on osteosarcoma tumor tissue (OTA)} in terms of balanced accuracy, ROC AUC, and weighted F1 score. The best-performing model for each metric is bolded, with 95\% CI in parentheses.}
\label{tab:ota_full}
\end{table}

\begin{table}
\centering
\begin{tabular}{lllll}
\toprule
Model &  Adaptation &  Balanced accuracy &  ROC AUC    & Weighted F1\\
\midrule
CONCH & PathFiT Disable & 91.43 (90.9-91.95) & 99.98 (99.98-99.98) & 98.45 (98.36-98.55) \\
      & PathFiT Enable & \textbf{95.24 (93.8-96.68)} & \textbf{99.99 (99.99-99.99)} & \textbf{99.17 (99.05-99.29)}\\
\midrule
UNI   & PathFiT Disable & 97.32 (96.88-97.76) & \textbf{99.99 (99.99-99.99)} & \textbf{99.21 (99.15-99.27)}\\
      & PathFiT Enable & \textbf{98.02 (96.73-99.3)} & \textbf{99.99 (99.99-99.99)} & 99.18 (99.04-99.32)\\
\bottomrule
\end{tabular}
\caption{\textbf{ROI-level supervised classification results on ESCA tissue (TolkachData)} in terms of balanced accuracy, ROC AUC, and weighted F1 score. The best-performing model for each metric is bolded, with 95\% CI in parentheses.}
\label{tab:tolkachdata_full}
\end{table}

\begin{table}
\centering
\begin{tabular}{lllll}
\toprule
Model &  Adaptation &  Balanced accuracy &  ROC AUC    & Weighted F1\\
\midrule
CONCH & PathFiT Disable & 82.91 (82.21-83.61) & 93.00 (92.75-93.26) & 86.65 (86.30-87.00) \\
      & PathFiT Enable & \textbf{85.85 (85.08-86.62)} & \textbf{93.81 (93.21-94.40)} & \textbf{86.94 (85.57-88.32)}\\
\midrule
UNI   & PathFiT Disable & 82.87 (82.29-83.45) & 93.16 (92.83-93.50) & 86.88 (86.19-87.56)\\
      & PathFiT Enable & \textbf{87.28 (86.49-88.06)} & \textbf{94.65 (94.21-95.10)} & \textbf{88.20 (86.59-89.81)}\\
\bottomrule
\end{tabular}
\caption{\textbf{ROI-level supervised classification results on colorectal precancer detection (MHIST)} in terms of balanced accuracy, ROC AUC, and weighted F1 score. The best-performing model for each metric is bolded, with 95\% CI in parentheses.}
\label{tab:mhist_full}
\end{table}

\begin{table}
\centering
\begin{tabular}{lllll}
\toprule
Model &  Adaptation &  Dice &  Precision    & Recall\\
\midrule
CONCH & PathFiT Disable & \textbf{80.28 (80.12-80.44)} & 78.14 (76.22-80.06) & \textbf{83.29 (81.05-85.53)}
 \\
      & PathFiT Enable & 80.24 (79.94-80.55) & \textbf{80.25 (79.42-81.07)}  & 81.02 (79.86-82.18)
\\
\midrule
UNI   & PathFiT Disable & 81.96 (81.69-82.22) & 80.06 (79.38-80.74)  & 84.50 (83.44-85.57) \\
      & PathFiT Enable & \textbf{82.18 (81.73-82.62)} & \textbf{80.27 (78.87-81.68)}  & \textbf{84.91 (82.52-87.30)}
\\
\bottomrule
\end{tabular}
\caption{\textbf{ROI-level supervised segmentation results on epithelial cell (SegPath)} in terms of dice, precision, and recall. The best-performing model for each metric is bolded, with 95\% CI in parentheses.}
\label{tab:panCK}
\end{table}

\begin{table}
\centering
\begin{tabular}{lllll}
\toprule
Model &  Adaptation &  Dice &  Precision    & Recall\\
\midrule
CONCH & PathFiT Disable & 89.37 (89.15-89.59) & \textbf{88.55 (86.84-90.26)}	 & 90.35 (88.49-92.20)
 \\
      & PathFiT Enable & \textbf{90.00 (89.87-90.12)} & 88.24 (87.38-89.10)	 & \textbf{91.93 (90.99-92.87)}

\\
\midrule
UNI   & PathFiT Disable & 91.05 (90.98-91.12) & 89.85 (89.07-90.63)	 & \textbf{92.34 (91.45-93.24)} \\
      & PathFiT Enable & \textbf{91.53 (91.34-91.72)} & \textbf{91.31 (90.81-91.81)}  & 91.80 (90.99-92.61)
\\
\bottomrule
\end{tabular}
\caption{\textbf{ROI-level supervised segmentation results on colon gland (Warwick-QU)} in terms of dice, precision, and recall. The best-performing model for each metric is bolded, with 95\% CI in parentheses.}
\label{tab:warwick-qu}
\end{table}

\begin{table}
\centering
\begin{tabular}{lllll}
\toprule
Model &  Adaptation &  Dice &  Precision    & Recall\\
\midrule
CONCH & PathFiT Disable & 63.41 (63.01-63.81) & \textbf{67.08 (65.61-68.55)}	 & 62.58 (62.07-63.09) \\
      & PathFiT Enable & \textbf{64.58 (64.19-64.96)} & 65.52 (64.06-66.97)	 & \textbf{65.62 (64.78-66.46)} \\
\midrule
UNI   & PathFiT Disable & 64.97 (64.38-65.56) & 68.00 (66.45-69.54)  & 	64.49 (62.71-66.28) \\
      & PathFiT Enable & \textbf{66.11 (65.37-66.85)} & \textbf{68.33 (66.47-70.19)}	 & \textbf{66.27 (64.97-67.57}) \\
\bottomrule
\end{tabular}
\caption{\textbf{ROI-level supervised segmentation results on colon nuclei identification (CoNIC)} in terms of dice, precision, and recall. The best-performing model for each metric is bolded, with 95\% CI in parentheses.}
\label{tab:conic}
\end{table}

\begin{table}
\centering
\begin{tabular}{lllll}
\toprule
Model &  Adaptation &  Balanced accuracy &  ROC AUC    & Weighted F1\\
\midrule
CONCH & PathFiT Disable & 82.41 (81.62-83.20) & 99.44 (99.42-99.46) & \textbf{88.30 (88.03-88.56)} \\
      & PathFiT Enable & \textbf{83.93 (83.39-84.47)} & \textbf{99.49 (99.46-99.51)}	 & 88.30 \textbf{(88.08-88.52)} \\
\midrule
UNI   & PathFiT Disable & 83.81 (83.13-84.48) & \textbf{99.33 (99.24-99.42)}	 & 88.36 (87.84-88.88) \\
      & PathFiT Enable & \textbf{85.02 (84.34-85.70)} & 99.30 (99.22-99.37)	 & \textbf{89.07 (88.41-89.73)}
 \\
\bottomrule
\end{tabular}
\caption{\textbf{Slide-level supervised results on OncoTree classification (TCGA)} in terms of balanced accuracy, ROC AUC, and weighted F1 score. The best-performing model for each metric is bolded, with 95\% CI in parentheses.}
\label{tab:oncotree}
\end{table}

\begin{table}
\centering
\begin{tabular}{lllll}
\toprule
Model &  Adaptation &  Balanced accuracy &  ROC AUC    & Weighted F1\\
\midrule
CONCH & PathFiT Disable & 91.41 (91.26-91.57) & \textbf{99.66 (99.65-99.67)}	 & 90.86 (90.45-91.26) \\
      & PathFiT Enable & \textbf{92.08 (91.76-92.39)} & 99.64 (99.63-99.65)	 & \textbf{91.52 (90.99-92.05)}
 \\
\midrule
UNI   & PathFiT Disable & \textbf{90.16 (89.88-90.44)} & \textbf{99.49 (99.43-99.55)}	 & \textbf{89.87 (89.30-90.43)} \\
      & PathFiT Enable & 89.41 (88.60-90.22) & 99.36 (99.33-99.39)	 & 88.81 (87.65-89.98)\\
\bottomrule
\end{tabular}
\caption{\textbf{Slide-level supervised results on pan-cancer classification (CPTAC)} in terms of balanced accuracy, ROC AUC, and weighted F1 score. The best-performing model for each metric is bolded, with 95\% CI in parentheses.}
\label{tab:cptac}
\end{table}

\begin{table}
\centering
\begin{tabular}{lllll}
\toprule
Model &  Adaptation &  Balanced accuracy &  ROC AUC    & Weighted F1\\
\midrule
CONCH & PathFiT Disable & 62.15 (59.27-65.03) & \textbf{91.87 (90.67-93.06)}	 & 83.21 (82.68-83.74) \\
      & PathFiT Enable & \textbf{63.02 (59.14-66.90)} & 90.20 (89.49-90.91)	 & \textbf{84.10 (83.03-85.17)} \\
\midrule
UNI   & PathFiT Disable & 62.29 (58.68-65.91) & \textbf{91.18 (90.42-91.94)}	 & 83.37 (82.94-83.80) \\
      & PathFiT Enable & \textbf{66.29 (65.47-67.10)} & 90.47 (89.36-91.58)	 & \textbf{85.44 (85.05-85.83)}\\
\bottomrule
\end{tabular}
\caption{\textbf{Slide-level supervised results on breast metastasis fine-grained detection (Camelyon+)} in terms of balanced accuracy, ROC AUC, and weighted F1 score. The best-performing model for each metric is bolded, with 95\% CI in parentheses.}
\label{tab:camelyon}
\end{table}

\begin{table}
\centering
\begin{tabular}{lllll}
\toprule
Model &  Adaptation &  Balanced accuracy &  ROC AUC    & Weighted F1\\
\midrule
CONCH & PathFiT Disable & 70.68 (69.89-71.46) & 90.44 (90.04-90.85)	 & 68.82 (67.62-70.02) \\
      & PathFiT Enable & \textbf{72.83 (72.03-73.62)} & \textbf{91.41 (91.25-91.57)}	 & \textbf{71.61 (71.54-71.69)} \\
\midrule
UNI   & PathFiT Disable & 72.01 (70.87-73.15) & 91.04 (90.59-91.48)	 & 71.47 (71.07-71.86) \\
      & PathFiT Enable & \textbf{75.50 (74.59-76.40)} & \textbf{92.28 (92.20-92.37)}	 & \textbf{74.72 (74.16-75.28)}\\
\bottomrule
\end{tabular}
\caption{\textbf{Slide-level supervised results on cervical lesions detection (TissueNet)} in terms of balanced accuracy, ROC AUC, and weighted F1 score. The best-performing model for each metric is bolded, with 95\% CI in parentheses.}
\label{tab:tissuenet}
\end{table}

\begin{table}
\centering
\begin{tabular}{lllll}
\toprule
Model &  Adaptation &  Balanced accuracy &  ROC AUC    & Weighted F1\\
\midrule
CONCH & PathFiT Disable & 68.97 (67.23-70.72) & 97.68 (97.58-97.77)	 & 73.47 (72.32-74.62) \\
      & PathFiT Enable & \textbf{70.05 (68.90-71.19)} & \textbf{97.71 (97.42-98.00)}	 & \textbf{75.27 (74.86-75.69)}\\
\midrule
UNI   & PathFiT Disable & 69.19 (68.52-69.87) & \textbf{97.55 (97.22-97.88)}	 & 74.89 (74.27-75.52) \\
      & PathFiT Enable & \textbf{71.26 (70.88-71.63)} & 97.53 (97.44-97.62)	 & \textbf{76.56 (76.22-76.89)} \\
\bottomrule
\end{tabular}
\caption{\textbf{Slide-level supervised results on brain tumor subtyping (EBRAINS)} in terms of balanced accuracy, ROC AUC, and weighted F1 score. The best-performing model for each metric is bolded, with 95\% CI in parentheses.}
\label{tab:brain_subtyping}
\end{table}

\begin{table}
\centering
\begin{tabular}{lllll}
\toprule
Model &  Adaptation &  Balanced accuracy &  ROC AUC    & Weighted F1\\
\midrule
CONCH & PathFiT Disable & 60.85 (60.50-61.19) & \textbf{90.09 (89.24-90.94)}	 & 79.61 (78.87-80.36) \\
      & PathFiT Enable & \textbf{64.15 (63.05-65.26)} & 90.08 (88.03-92.14)	 & \textbf{80.96 (78.93-83.00)}\\
\midrule
UNI   & PathFiT Disable & 52.15 (47.16-57.14) & 87.22 (84.89-89.56)	 & 77.18 (73.53-80.82) \\
      & PathFiT Enable & \textbf{62.58 (59.80-65.35)} & \textbf{90.17 (89.09-91.26)}	 & \textbf{80.19 (78.93-81.45)}
 \\
\bottomrule
\end{tabular}
\caption{\textbf{Slide-level supervised results on glioma histomolecular subtyping (EBRAINS)} in terms of balanced accuracy, ROC AUC, and weighted F1 score. The best-performing model for each metric is bolded, with 95\% CI in parentheses.}
\label{tab:glioma-molecular}
\end{table}

\begin{table}
\centering
\begin{tabular}{lllll}
\toprule
Model &  Adaptation &  Balanced accuracy &  ROC AUC    & Weighted F1\\
\midrule
CONCH & PathFiT Disable & 41.50 (39.05-43.96) & 82.34 (81.42-83.25)	 & 44.74 (40.21-49.27) \\
      & PathFiT Enable & \textbf{44.70 (42.00-47.39)} & \textbf{83.68 (81.94-85.42)}	 & \textbf{48.03 (45.99-50.08)}
\\
\midrule
UNI   & PathFiT Disable & 36.83 (36.25-37.41) & 78.20 (75.83-80.56)	 & 41.84 (41.17-42.51) \\
      & PathFiT Enable & \textbf{38.83 (37.81-39.85)} & \textbf{82.00 (79.73-84.28)}	 & \textbf{42.67 (39.38-45.95)} \\
\bottomrule
\end{tabular}
\caption{\textbf{Slide-level supervised results on BRCA fine-grained subtyping (BRACS)} in terms of balanced accuracy, ROC AUC, and weighted F1 score. The best-performing model for each metric is bolded, with 95\% CI in parentheses.}
\label{tab:bracs-f}
\end{table}

\begin{table}
\centering
\begin{tabular}{lllll}
\toprule
Model &  Adaptation &  Balanced accuracy &  ROC AUC    & Weighted F1\\
\midrule
CONCH & PathFiT Disable & 69.64 (68.37-70.91) & \textbf{89.20 (86.77-91.62)}	 & 71.05 (69.82-72.27) \\
      & PathFiT Enable & \textbf{72.74 (71.50-73.97)} & 88.35 (86.96-89.74)	 & \textbf{74.56 (73.53-75.59)} \\
\midrule
UNI   & PathFiT Disable & 67.53 (63.53-71.52) & 86.43 (84.22-88.64)	 & 68.27 (65.26-71.28) \\
      & PathFiT Enable & \textbf{74.85 (74.38-75.31)} & \textbf{89.59 (88.34-90.84)}	 & \textbf{75.42 (74.72-76.12)} \\
\bottomrule
\end{tabular}
\caption{\textbf{Slide-level supervised results on BRCA coarsed-grained subtyping (BRACS)} in terms of balanced accuracy, ROC AUC, and weighted F1 score. The best-performing model for each metric is bolded, with 95\% CI in parentheses.}
\label{tab:bracs-c}
\end{table}

\begin{table}
\centering
\begin{tabular}{lllll}
\toprule
Model &  Adaptation &  Balanced accuracy &  ROC AUC    & Weighted F1\\
\midrule
CONCH & PathFiT Disable & 85.28 (84.02-86.54) & 92.24 (91.43-93.05)	 & 87.16 (84.46-89.87) \\
      & PathFiT Enable & \textbf{85.59 (84.68-86.50)} & \textbf{92.56 (92.24-92.88)}	 & \textbf{89.48 (88.44-90.52)} \\
\midrule
UNI   & PathFiT Disable & 84.48 (82.34-86.62) & 91.12 (89.67-92.58)	 & \textbf{89.49 (87.90-91.09)} \\
      & PathFiT Enable & \textbf{86.49 (85.63-87.36)} & \textbf{92.02 (91.61-92.43)}	 & 89.16 (87.86-90.45) \\
\bottomrule
\end{tabular}
\caption{\textbf{Slide-level supervised results on glioma \textit{IHD1} prediction (EBRAINS)} in terms of balanced accuracy, ROC AUC, and weighted F1 score. The best-performing model for each metric is bolded, with 95\% CI in parentheses.}
\label{tab:idh1}
\end{table}

\begin{table}
\centering
\begin{tabular}{lllll}
\toprule
Model &  Adaptation &  Balanced accuracy &  ROC AUC    & Weighted F1\\
\midrule
CONCH & PathFiT Disable & 76.67 (72.91-80.43) & 86.39 (85.88-86.90)	 & 73.99 (68.36-79.63) \\
      & PathFiT Enable & \textbf{79.08 (77.62-80.54)} & \textbf{86.53 (84.48-88.59)}	 & \textbf{77.15 (75.34-78.97)} \\
\midrule
UNI   & PathFiT Disable & 69.79 (65.95-73.63) & 77.58 (74.69-80.47)	 & 68.89 (64.40-73.39) \\
      & PathFiT Enable & \textbf{74.86 (73.60-76.13}) & \textbf{83.95 (83.17-84.73)}	 & \textbf{75.47 (74.05-76.88)} \\
\bottomrule
\end{tabular}
\caption{\textbf{Slide-level supervised results on BRCA \textit{HER2} prediction (HEROHE)} in terms of balanced accuracy, ROC AUC, and weighted F1 score. The best-performing model for each metric is bolded, with 95\% CI in parentheses.}
\label{tab:her2}
\end{table}

\begin{table}
\centering
\begin{tabular}{lllll}
\toprule
Model &  Adaptation &  Balanced accuracy &  ROC AUC    & Weighted F1\\
\midrule
CONCH & PathFiT Disable & 35.37 (32.18-38.56) & \textbf{70.73 (69.13-72.34)	} & 51.55 (49.77-53.33) \\
      & PathFiT Enable & \textbf{37.90 (36.56-39.24)} & 69.32 (67.46-71.19)	 & \textbf{53.38 (51.61-55.14)} \\
\midrule
UNI   & PathFiT Disable & \textbf{34.90 (30.49-39.31)} & 66.18 (62.63-69.72)	 & \textbf{51.79 (48.54-55.05)} \\
      & PathFiT Enable & 33.94 (32.05-35.83) & \textbf{66.86 (64.08-69.65)}	 & 51.06 (49.73-52.38) \\
\bottomrule
\end{tabular}
\caption{\textbf{Slide-level supervised results on BRCA IHC scoring (HEROHE)} in terms of balanced accuracy, ROC AUC, and weighted F1 score. The best-performing model for each metric is bolded, with 95\% CI in parentheses.}
\label{tab:ihc}
\end{table}

\begin{table}
\centering
\begin{tabular}{lllll}
\toprule
Model &  Adaptation &  Balanced accuracy &  ROC AUC    & Weighted F1\\
\midrule
CONCH & PathFiT Disable & 50.46 (49.81-51.11) & 83.86 (83.11-84.61)	 & 54.38 (52.62-56.14) \\
      & PathFiT Enable & \textbf{51.13 (49.05-53.21)} & \textbf{83.95 (83.79-84.12)}	 & \textbf{56.09 (55.71-56.47)}
 \\
\midrule
UNI   & PathFiT Disable & \textbf{46.37 (45.44-47.31)} & 82.68 (82.47-82.88)	 & \textbf{55.31 (55.05-55.57)}\\
      & PathFiT Enable & 46.21 (45.66-46.77) & \textbf{82.99 (82.6-83.37)}	 & 53.48 (53.40-53.56) \\
\bottomrule
\end{tabular}
\caption{\textbf{Slide-level supervised results on pan-cancer TILs scoring (TCGA)} in terms of balanced accuracy, ROC AUC, and weighted F1 score. The best-performing model for each metric is bolded, with 95\% CI in parentheses.}
\label{tab:tils}
\end{table}

\begin{table}
\centering
\begin{tabular}{lllll}
\toprule
Model &  Adaptation &  Balanced accuracy &  ROC AUC    & Weighted F1\\
\midrule
CONCH & PathFiT Disable & 90.51 (90.21-90.81) & 96.16 (95.91-96.41)	 & 89.89 (89.80-89.98) \\
      & PathFiT Enable & \textbf{93.69 (93.33-94.05)} & \textbf{97.79 (97.75-97.82)}	 & \textbf{93.24 (92.82-93.67)} \\
\midrule
UNI   & PathFiT Disable & 88.90 (88.68-89.12) & 95.18 (94.99-95.37)	 & 88.39 (88.00-88.79)\\
      & PathFiT Enable & \textbf{93.96 (93.42-94.50)} & \textbf{98.17 (98.01-98.34)}	 & \textbf{93.70 (93.22-94.18)} \\
\bottomrule
\end{tabular}
\caption{\textbf{Slide-level supervised results on PRAD screening (PANDA: Karolinska)} in terms of balanced accuracy, ROC AUC, and weighted F1 score. The best-performing model for each metric is bolded, with 95\% CI in parentheses.}
\label{tab:panda-karo-2}
\end{table}

\begin{table}
\centering
\begin{tabular}{lllll}
\toprule
Model &  Adaptation &  Balanced accuracy &  ROC AUC    & Weighted F1\\
\midrule
CONCH & PathFiT Disable & 53.00 (51.36-54.63) & 90.17 (89.83-90.5)	 & 66.49 (65.15-67.83) \\
      & PathFiT Enable & \textbf{56.88 (56.13-57.63)} & \textbf{91.68 (91.38-91.98)}	 & \textbf{71.11 (70.37-71.85)} \\
\midrule
UNI   & PathFiT Disable & 52.24 (51.07-53.41) & 87.88 (87.33-88.42)	 & 63.99 (63.07-64.91)\\
      & PathFiT Enable & \textbf{59.34 (56.21-62.47)} & \textbf{92.46 (91.90-93.02)}	 & \textbf{71.75 (69.99-73.51)}
 \\
\bottomrule
\end{tabular}
\caption{\textbf{Slide-level supervised results on PRAD grading (PANDA: Karolinska)} in terms of balanced accuracy, ROC AUC, and weighted F1 score. The best-performing model for each metric is bolded, with 95\% CI in parentheses.}
\label{tab:panda-karo-6}
\end{table}

\begin{table}
\centering
\begin{tabular}{lllll}
\toprule
Model &  Adaptation &  Balanced accuracy &  ROC AUC    & Weighted F1\\
\midrule
CONCH & PathFiT Disable & 92.57 (91.95-93.18) & 96.78 (96.59-96.97)	 & 91.90 (91.25-92.54) \\
      & PathFiT Enable & \textbf{92.77 (92.38-93.16)} & \textbf{97.00 (96.84-97.17)}	 & \textbf{92.01 (91.43-92.58)} \\
\midrule
UNI   & PathFiT Disable & 92.19 (91.87-92.52) & 96.69 (96.56-96.83)	 & 91.36 (90.96-91.75)\\
      & PathFiT Enable & \textbf{93.74 (93.26-94.22)} & \textbf{97.33 (97.12-97.55)}	 & \textbf{92.94 (92.08-93.80)} \\
\bottomrule
\end{tabular}
\caption{\textbf{Slide-level supervised results on PRAD screening (PANDA: Radboud)} in terms of balanced accuracy, ROC AUC, and weighted F1 score. The best-performing model for each metric is bolded, with 95\% CI in parentheses.}
\label{tab:panda-rad-2}
\end{table}

\begin{table}
\centering
\begin{tabular}{lllll}
\toprule
Model &  Adaptation &  Balanced accuracy &  ROC AUC    & Weighted F1\\
\midrule
CONCH & PathFiT Disable & 58.04 (56.96-59.12) & 88.26 (87.91-88.61)	 & 57.21 (55.90-58.53) \\
      & PathFiT Enable & \textbf{60.92 (60.12-61.71)} & \textbf{88.68 (88.35-89.01)}	 & \textbf{61.27 (60.11-62.43)} \\
\midrule
UNI   & PathFiT Disable & 59.49 (58.35-60.63) & 88.69 (88.46-88.92)	 & 59.75 (58.88-60.61)\\
      & PathFiT Enable & \textbf{63.11 (61.81-64.41)} & \textbf{89.64 (89.22-90.06)}	 & \textbf{63.05 (61.35-64.76)} \\
\bottomrule
\end{tabular}
\caption{\textbf{Slide-level supervised results on PRAD grading (PANDA: Radboud)} in terms of balanced accuracy, ROC AUC, and weighted F1 score. The best-performing model for each metric is bolded, with 95\% CI in parentheses.}
\label{tab:panda-rad-6}
\end{table}

\begin{table}
\centering
\begin{tabular}{llll}
\toprule
Adaptation &  Balanced accuracy &  ROC AUC    & Weighted F1\\
\midrule
PathFiT Disable & 71.93 (70.85-73.01) & 90.08 (89.4-90.77)	 & 75.54 (74.97-76.11) \\
PathFiT Enable & \textbf{76.35 (72.77-79.94)} & \textbf{91.81 (90.93-92.7)}	 & \textbf{79.79 (76.62-82.96)} \\
\bottomrule
\end{tabular}
\caption{\textbf{Slide-level supervised results of CONCH on in-house cervical inflammatory tissue classification (XJH)} in terms of balanced accuracy, ROC AUC, and weighted F1 score. The best-performing model for each metric is bolded, with 95\% CI in parentheses.}
\label{tab:xj-cervical}
\end{table}

\begin{table}
\centering
\begin{tabular}{lllll}
\toprule
Model &  Adaptation &  Balanced accuracy &  ROC AUC    & Weighted F1\\
\midrule
CONCH & PathFiT Disable & 26.92 (25.8-28.05) & 76.69 (76.35-77.03)	 & 24.19 (22.76-25.62) \\
      & PathFiT Enable & \textbf{38.24 (35.7-40.78)} & \textbf{84.84 (84.33-85.35)}	 & \textbf{37.79 (35.39-40.18)} \\
\midrule
UNI   & PathFiT Disable & 28.99 (28.03-29.96) & 78.14 (77.69-78.58)	 & 25.17 (24.23-26.11)\\
      & PathFiT Enable & \textbf{40.23 (39.02-41.44)} & \textbf{85.05 (84.28-85.82)}	 & \textbf{40.43 (39.25-41.61)}\\
\bottomrule
\end{tabular}
\caption{\textbf{ROI-level supervised results on in-house glomerular structure classification of transmission electron microscopy (XJH)} in terms of balanced accuracy, ROC AUC, and weighted F1 score. The best-performing model for each metric is bolded, with 95\% CI in parentheses.}
\label{tab:em_stain}
\end{table}

\begin{table}
\centering
\begin{tabular}{lllll}
\toprule
Model &  Adaptation &  Balanced accuracy &  ROC AUC    & Weighted F1\\
\midrule
CONCH & PathFiT Disable & 44.08 (42.59-45.57) & 74.97 (72.76-77.19)	 & 50.24 (48.41-52.07) \\
      & PathFiT Enable & \textbf{60.50 (58.10-62.91)} & \textbf{87.45 (86.68-88.23)}	 & \textbf{66.09 (64.38-67.81)}\\
\midrule
UNI   & PathFiT Disable & 47.41 (43.90-50.92) & 75.70 (73.13-78.28)	 & 54.19 (50.71-57.66)\\
      & PathFiT Enable & \textbf{63.04 (60.02-66.06)} & \textbf{88.91 (87.59-90.24)}	 & \textbf{66.90 (62.62-71.18)}\\
\bottomrule
\end{tabular}
\caption{\textbf{ROI-level supervised results on in-house Masson's Trichrome glomerular classification (XJH)} in terms of balanced accuracy, ROC AUC, and weighted F1 score. The best-performing model for each metric is bolded, with 95\% CI in parentheses.}
\label{tab:masson}
\end{table}

\begin{table}
\centering
\begin{tabular}{lllll}
\toprule
Model &  Adaptation &  Balanced accuracy &  ROC AUC    & Weighted F1\\
\midrule
CONCH & PathFiT Disable & 59.16 (58.22-60.11) & 85.88 (85.52-86.23)	 & 63.80 (62.97-64.63) \\
      & PathFiT Enable & \textbf{72.41 (70.82-74.00)} & \textbf{92.18 (91.86-92.50)}	 & \textbf{73.35 (72.50-74.20)}\\
\midrule
UNI   & PathFiT Disable & 63.23 (62.17-64.28) & 87.34 (87.04-87.65)	 & 65.80 (64.99-66.61)\\
      & PathFiT Enable & \textbf{72.04 (70.55-73.53)} & \textbf{92.28 (91.64-92.92)}	 & \textbf{72.96 (71.40-74.53)}\\
\bottomrule
\end{tabular}
\caption{\textbf{ROI-level supervised results on in-house Periodic Acid-Schiff glomerular classification (XJH)} in terms of balanced accuracy, ROC AUC, and weighted F1 score. The best-performing model for each metric is bolded, with 95\% CI in parentheses.}
\label{tab:pas}
\end{table}

\begin{table}
\centering
\begin{tabular}{lllll}
\toprule
Model &  Adaptation &  Balanced accuracy &  ROC AUC    & Weighted F1\\
\midrule
CONCH & PathFiT Disable & 43.17 (38.94-47.41) & 75.79 (74.69-76.89)	 & 49.19 (44.84-53.54) \\
      & PathFiT Enable & \textbf{54.10 (51.89-56.32)} & \textbf{82.42 (81.63-83.22)}	 & \textbf{58.61 (56.77-60.45)}\\
\midrule
UNI   & PathFiT Disable & 41.22 (39.16-43.29) & 75.79 (74.44-77.15)	 & 47.92 (46.01-49.84)\\
      & PathFiT Enable & \textbf{57.37 (56.74-58.00)} & \textbf{84.86 (83.84-85.88)} & \textbf{61.16 (59.83-62.50)}\\
\bottomrule
\end{tabular}
\caption{\textbf{ROI-level supervised results on in-house Periodic Acid-Schiff Methenamine glomerular classification (XJH)} in terms of balanced accuracy, ROC AUC, and weighted F1 score. The best-performing model for each metric is bolded, with 95\% CI in parentheses.}
\label{tab:pasm}
\end{table}

\begin{table}
\centering
\begin{tabular}{lllll}
\toprule
Model &  Adaptation &  Balanced accuracy &  ROC AUC    & Weighted F1\\
\midrule
CONCH & PathFiT Disable & 77.65 (77.07-78.23) & 87.34 (86.86-87.82)	 & 79.47 (78.76-80.17)) \\
      & PathFiT Enable & \textbf{89.53 (88.48-90.57)} & \textbf{96.20 (95.77-96.63)}	 & \textbf{89.90 (89.07-90.73)}\\
\midrule
UNI   & PathFiT Disable & 87.15 (85.88-88.43) & 94.62 (93.91-95.33)	 & 87.51 (86.43-88.59)\\
      & PathFiT Enable & \textbf{92.83 (92.39-93.26)} & \textbf{98.09 (97.87-98.31)} & \textbf{92.83 (92.35-93.31)}
\\
\bottomrule
\end{tabular}
\caption{\textbf{ROI-level supervised results on in-house immunofluorescence sediment organization classification (XJH)} in terms of balanced accuracy, ROC AUC, and weighted F1 score. The best-performing model for each metric is bolded, with 95\% CI in parentheses.}
\label{tab:if_so}
\end{table}

\begin{table}
\centering
\begin{tabular}{lllll}
\toprule
Model &  Adaptation &  Balanced accuracy &  ROC AUC    & Weighted F1\\
\midrule
CONCH & PathFiT Disable & 76.84 (75.03-78.64) & 86.11 (85.30-86.91)	 & 76.90 (75.41-78.40) \\
      & PathFiT Enable & \textbf{90.57 (90.30-90.85)} & \textbf{96.93 (96.65-97.21)} & \textbf{90.66 (90.40-90.93)}\\
\midrule
UNI   & PathFiT Disable & 88.57 (87.87-89.28) & 96.28 (95.84-96.72)	 & 88.82 (88.30-89.33)\\
      & PathFiT Enable & \textbf{92.63 (91.71-93.56)} & \textbf{97.92 (97.71-98.13)} & \textbf{92.65 (91.72-93.58)}\\
\bottomrule
\end{tabular}
\caption{\textbf{ROI-level supervised results on in-house immunofluorescence deposit distribution detection (XJH)} in terms of balanced accuracy, ROC AUC, and weighted F1 score. The best-performing model for each metric is bolded, with 95\% CI in parentheses.}
\label{tab:if_dd}
\end{table}

\begin{table}
\centering
\begin{tabular}{lllll}
\toprule
Model &  Adaptation &  Balanced accuracy &  ROC AUC    & Weighted F1\\
\midrule
CONCH & PathFiT Disable & 79.76 (79.59-79.94) & 97.09 (97.05-97.14)	 & 81.88 (81.61-82.14) \\
      & PathFiT Enable & \textbf{82.25 (81.79-82.71)} & \textbf{97.48 (97.40-97.55)}	 & \textbf{83.88 (83.52-84.24)}\\
\midrule
UNI   & PathFiT Disable & 81.33 (81.09-81.57) & 97.36 (97.32-97.40)	 & 83.06 (82.73-83.39)\\
      & PathFiT Enable & \textbf{82.52 (82.14-82.89)} & \textbf{97.57 (97.38-97.76)} & \textbf{84.50 (84.27-84.73)}\\
\bottomrule
\end{tabular}
\caption{\textbf{ROI-level supervised results on immunohistochemistry tissue classification (MIHIC)} in terms of balanced accuracy, ROC AUC, and weighted F1 score. The best-performing model for each metric is bolded, with 95\% CI in parentheses.}
\label{tab:mihic}
\end{table}

\begin{table}
\centering
\begin{tabular}{llllll}
\toprule
Model & Resolution & Adaptation &  Balanced accuracy &  ROC AUC    & Weighted F1\\
\midrule
CONCH & 1024$\times$1024 & PathFiT Disable  & 78.50 (75.65-81.35) & 94.84 (93.79-95.89) & 78.07 (75.15-80.98) \\
 & & PathFiT Enable & \textbf{86.50 (85.09-87.91)} & \textbf{97.70 (97.18-98.22)} & \textbf{86.43 (84.98-87.87)} \\
 \midrule
 CONCH & 768$\times$768 & PathFiT Disable & 80.17 (76.19-84.15) & 95.86 (94.36-97.35) & 79.87 (76.04-83.71)\\
 & & PathFiT Enable & \textbf{88.33 (87.18-89.49)} & \textbf{98.17 (97.72-98.61)}	 & \textbf{88.34 (87.15-89.52)} \\
 \midrule
  CONCH & 512$\times$612 & PathFiT Disable & 84.00 (81.11-86.89) & 96.55 (95.98-97.11) & 83.81 (80.93-86.69)\\
 & & PathFiT Enable & \textbf{87.00 (83.71-90.29)} & \textbf{97.99 (97.60-98.38)} & \textbf{86.97 (83.67-90.28)} \\
 \midrule
  CONCH & 256$\times$256 & PathFiT Disable & 81.83 (78.80-84.86) & 94.68 (94.05-95.31) & 81.52 (78.37-84.66)\\
 & & PathFiT Enable & \textbf{87.50 (86.23-88.77)} & \textbf{97.52 (96.90-98.14)} & \textbf{87.47 (86.04-88.89)} \\
\bottomrule
\end{tabular}
\caption{\textbf{ROI-level supervised results of CONCH with different resolutions on BRCA subtyping (BACH)} in terms of balanced accuracy, ROC AUC, and weighted F1 score. The best-performing model for each metric is bolded, with 95\% CI in parentheses.}
\label{tab:bach_diff_res_conch}
\end{table}

\begin{table}
\centering
\begin{tabular}{llllll}
\toprule
Model & Resolution & Adaptation &  Balanced accuracy &  ROC AUC    & Weighted F1\\
\midrule
UNI & 1024$\times$1024 & PathFiT Disable  & 85.67 (84.08-87.25) & 96.63 (96.01-97.26) & 85.45 (83.85-87.05) \\
 & & PathFiT Enable & \textbf{92.17 (90.94-93.39)} & \textbf{99.10 (98.93-99.28)}	 & \textbf{92.14 (90.97-93.32)}
 \\
 \midrule
 UNI & 768$\times$768 & PathFiT Disable & 82.83 (80.67-85.00) & 96.61 (96.02-97.21) & 82.49 (80.14-84.83)\\
 & & PathFiT Enable & \textbf{91.33 (89.50-93.17)} & \textbf{99.29 (99.08-99.51)}	 & \textbf{91.27 (89.39-93.15)} \\
 \midrule
  UNI & 512$\times$512 & PathFiT Disable & 81.17 (79.13-83.21) & 95.67 (94.20-97.14) & 80.93 (78.96-82.90)\\
 & & PathFiT Enable & \textbf{92.33 (91.25-93.42)} & \textbf{99.30 (98.97-99.62)}	 & \textbf{92.30 (91.27-93.32)}
 \\
 \midrule
  UNI & 256$\times$256 & PathFiT Disable & 82.67 (79.77-85.56) & 95.72 (94.61-96.83) & 82.76 (79.84-85.68)\\
 & & PathFiT Enable & \textbf{89.67 (88.24-91.09)} & \textbf{98.65 (98.20-99.10)}	 & \textbf{89.69 (88.26-91.12)} \\
 
\bottomrule
\end{tabular}
\caption{\textbf{ROI-level supervised results of UNI with different resolutions on BRCA subtyping (BACH)} in terms of balanced accuracy, ROC AUC, and weighted F1 score. The best-performing model for each metric is bolded, with 95\% CI in parentheses.}
\label{tab:bach_diff_res_uni}
\end{table}

\begin{table}
\centering
\begin{tabular}{llllll}
\toprule
Model & Resolution & Adaptation &  Balanced accuracy &  ROC AUC    & Weighted F1\\
\midrule
CONCH & 1024$\times$1024 & PathFiT Disable  & 57.52 (56.55-58.49) & 89.53 (89.37-89.70) & 55.79 (54.48-57.09) \\
 & & PathFiT Enable & \textbf{63.40 (61.54-65.26)} & \textbf{91.77 (91.04-92.50)}	 & \textbf{63.10 (61.19-65.01)} \\
 \midrule
 CONCH & 768$\times$768 & PathFiT Disable & 60.09 (59.45-60.73) & 90.13 (90.02-90.24) & 58.34 (57.80-58.88)\\
 & & PathFiT Enable & \textbf{63.83 (62.81-64.84)} & \textbf{91.59 (90.81-92.37)}	 & \textbf{63.74 (62.83-64.66)} \\
 \midrule
  CONCH & 512$\times$512 & PathFiT Disable & 62.21 (61.96-62.45) & 90.80 (90.64-90.96) & 60.99 (60.66-61.33)\\
 & & PathFiT Enable & \textbf{64.47 (63.67-65.26)} & \textbf{91.89 (91.44-92.34)}	 & \textbf{64.29 (63.52-65.06)} \\
 \midrule
  CONCH & 256$\times$256 & PathFiT Disable & 62.06 (61.19-62.92) & 90.45 (90.32-90.58) & 60.58 (59.49-61.66)\\
 & & PathFiT Enable & \textbf{64.74 (63.72-65.76)} & \textbf{91.39 (90.77-92.01)}	 & \textbf{64.45 (63.24-65.65)} \\
\bottomrule
\end{tabular}
\caption{\textbf{ROI-level supervised results of CONCH with different resolutions on BRCA fine-grained subtyping (BRACS)} in terms of balanced accuracy, ROC AUC, and weighted F1 score. The best-performing model for each metric is bolded, with 95\% CI in parentheses.}
\label{tab:bracs_diff_res_conch}
\end{table}

\begin{table}
\centering
\begin{tabular}{llllll}
\toprule
Model & Resolution & Adaptation &  Balanced accuracy &  ROC AUC    & Weighted F1\\
\midrule
UNI & 1024$\times$1024 & PathFiT Disable  & 59.30 (57.94-60.66) & 90.43 (90.06-90.81) & 58.30 (56.58-60.03)
 \\
 & & PathFiT Enable & \textbf{63.61 (61.70-65.53)} & \textbf{91.42 (90.89-91.95)}	 & \textbf{63.07 (61.06-65.07)}
 \\
 \midrule
 UNI & 768$\times$768 & PathFiT Disable & 61.27 (59.58-62.95) & 90.86 (90.60-91.13) & 60.14 (58.16-62.12)\\
 & & PathFiT Enable & \textbf{65.03 (63.06-67.01)} & \textbf{91.95 (91.32-92.58)}	 & \textbf{64.32 (62.60-66.04)} \\
 \midrule
  UNI & 512$\times$512 & PathFiT Disable & 62.29 (61.04-63.55) & 91.25 (90.94-91.55) & 62.04 (60.69-63.39)\\
 & & PathFiT Enable & \textbf{65.36 (64.30-66.42)} & \textbf{92.61 (92.38-92.83)} &\textbf{ 65.00 (64.00-65.99)}
 \\
 \midrule
  UNI & 256$\times$256 & PathFiT Disable & 60.63 (60.05-61.21) & 90.20 (90.00-90.39) & 60.89 (60.34-61.44)\\
 & & PathFiT Enable & \textbf{65.48 (64.38-66.58)} & \textbf{91.71 (91.07-92.34)}	 & \textbf{65.11 (64.16-66.06)}
 \\
 
\bottomrule
\end{tabular}
\caption{\textbf{ROI-level supervised results of UNI with different resolutions on BRCA fine-grained subtyping (BRACS)} in terms of balanced accuracy, ROC AUC, and weighted F1 score. The best-performing model for each metric is bolded, with 95\% CI in parentheses.}
\label{tab:bracs_diff_res_uni}
\end{table}

\begin{table}
\centering
\begin{tabular}{llllll}
\toprule
Model & Resolution & Adaptation &  Balanced accuracy &  ROC AUC    & Weighted F1\\
\midrule
CONCH & 1024$\times$1024 & PathFiT Disable  & 93.15 (92.28-94.02) & 99.10 (98.85-99.36) & 94.17 (93.39-94.95) \\
 & & PathFiT Enable & \textbf{95.37 (94.88-95.85)} & \textbf{99.48 (99.43-99.52)}	 & \textbf{95.67 (95.17-96.18)} \\
 \midrule
 CONCH & 768$\times$768 & PathFiT Disable & 93.33 (92.64-94.02) & 99.27 (99.16-99.38) & 94.47 (93.87-95.07)\\
 & & PathFiT Enable & \textbf{94.96 (94.62-95.30)} & \textbf{99.39 (99.29-99.49)}	 & \textbf{95.07 (94.55-95.59)} \\
 \midrule
  CONCH & 512$\times$512 & PathFiT Disable & 92.97 (92.07-93.86) & 99.22 (99.12-99.32) & 94.09 (93.35-94.82)\\
 & & PathFiT Enable & \textbf{94.65 (94.37-94.92)} & \textbf{99.40 (99.33-99.47)}	 & \textbf{95.24 (94.84-95.64)} \\
 \midrule
  CONCH & 256$\times$256 & PathFiT Disable & 93.44 (92.69-94.19) & 99.17 (99.02-99.32) & 94.42 (93.94-94.90)\\
 & & PathFiT Enable & \textbf{94.50 (94.19-94.81)} & \textbf{99.46 (99.38-99.53)}& \textbf{95.00 (94.85-95.14)} \\
\bottomrule
\end{tabular}
\caption{\textbf{ROI-level supervised results of CONCH with different resolutions on osteosarcoma tumor tissue classification (OTA)} in terms of balanced accuracy, ROC AUC, and weighted F1 score. The best-performing model for each metric is bolded, with 95\% CI in parentheses.}
\label{tab:ota_diff_res}
\end{table}

\begin{table}
\centering
\begin{tabular}{llllll}
\toprule
Model & Resolution & Adaptation &  Balanced accuracy &  ROC AUC    & Weighted F1\\
\midrule
UNI & 1024$\times$1024 & PathFiT Disable  & 93.40 (92.83-93.97) & 99.19 (99.08-99.31) & 94.72 (94.23-95.21)
 \\
 & & PathFiT Enable & \textbf{95.28 (94.54-96.01)} & \textbf{99.47 (99.42-99.52)} & \textbf{95.55 (94.83-96.27)}
 \\
 \midrule
 UNI & 768$\times$768 & PathFiT Disable & 92.93 (92.27-93.58) & 99.21 (99.16-99.26) & 94.23 (93.65-94.82)\\
 & & PathFiT Enable & \textbf{95.01 (94.48-95.53)} & \textbf{99.42 (99.36-99.49)}	 & \textbf{95.31 (94.83-95.79)}
 \\
 \midrule
  UNI & 512$\times$512 & PathFiT Disable & 92.32 (91.91-92.72) & 99.13 (99.08-99.17 & 93.51 (93.28-93.74)\\
 & & PathFiT Enable & \textbf{94.36 (93.73-94.99)} & \textbf{99.35 (99.28-99.42)}	 & \textbf{94.76 (94.13-95.39)}
 \\
 \midrule
  UNI & 256$\times$256 & PathFiT Disable & 93.21 (92.57-93.84) & 99.25 (99.20-99.29) & 94.06 (93.44-94.67)
\\
 & & PathFiT Enable & \textbf{95.21 (94.23-96.20)} & \textbf{99.45 (99.39-99.51)}	 & \textbf{95.61 (94.87-96.34)}
 \\
 
\bottomrule
\end{tabular}
\caption{\textbf{ROI-level supervised results of UNI with different resolutions on osteosarcoma tumor tissue classification (OTA)} in terms of balanced accuracy, ROC AUC, and weighted F1 score. The best-performing model for each metric is bolded, with 95\% CI in parentheses.}
\label{tab:ota_diff_res}
\end{table}

\begin{table}
\centering
\begin{tabular}{lllll}
\toprule
Shot number &  Adaptation &  Balanced accuracy &  ROC AUC    & Weighted F1\\
\midrule
1-shot & PathFiT Disable & 12.71 (11.94-13.49) & 70.94 (69.95-71.93) & 14.47 (12.83-16.11) \\
      & PathFiT Enable & \textbf{24.74 (23.14-26.34)} & \textbf{82.83 (81.95-83.71)} & \textbf{27.61 (25.52-29.71)}\\
\midrule
2-shot & PathFiT Disable & 23.18 (22.28-24.08) & 79.94 (79.23-80.64) & 24.48 (22.76-26.20) \\
      & PathFiT Enable & \textbf{32.72 (30.28-35.15)} & \textbf{86.13 (84.78-87.48)} & \textbf{32.15 (27.92-36.38)}\\
\midrule
4-shot & PathFiT Disable & 36.84 (36.24-37.45) & 87.49 (87.31-87.68) & 37.66 (37.11-38.21) \\
      & PathFiT Enable & \textbf{45.63 (44.48-46.79)} & \textbf{91.17 (90.73-91.62)} & \textbf{46.44 (45.19-47.69)}\\
\midrule
8-shot & PathFiT Disable & 48.93 (48.45-49.41) & 92.43 (92.34-92.52) & 50.04 (49.41-50.66) \\
      & PathFiT Enable & \textbf{55.31 (54.22-56.39)} & \textbf{93.92 (93.56-94.28)} & \textbf{56.45 (55.12-57.78)}\\
\midrule
16-shot & PathFiT Disable & \textbf{58.16 (57.83-58.48)} & \textbf{95.08 (95.01-95.16)} & \textbf{59.36 (58.89-59.83)} \\
      & PathFiT Enable & 57.09 (56.31-57.86) & 94.94 (94.78-95.11) & 57.19 (56.05-58.33)\\
\bottomrule
\end{tabular}
\caption{\textbf{Few-shot results of CONCH using text prompt learning on pan-cancer classification (TCGA)} in terms of balanced accuracy, ROC AUC, and weighted F1 score. The best-performing model for each metric is bolded, with 95\% CI in parentheses.}
\label{tab:few-shot-tcga-conch}
\end{table}

\begin{table}
\centering
\begin{tabular}{lllll}
\toprule
Shot number &  Adaptation &  Balanced accuracy &  ROC AUC    & Weighted F1\\
\midrule
1-shot & PathFiT Disable & 19.19 (18.25-20.13) & 75.52 (74.16-76.88) & 21.28 (19.60-22.97) \\
      & PathFiT Enable & \textbf{28.01 (26.74-29.29)} & \textbf{82.01 (80.64-83.39)} & \textbf{30.44 (28.69-32.18)}\\
\midrule
2-shot & PathFiT Disable & 29.81 (29.62-29.99) & 82.74 (82.03-83.45) & 31.54 (30.92-32.16) \\
      & PathFiT Enable & \textbf{35.19 (33.48-36.91)} & \textbf{86.36 (85.63-87.09)} & \textbf{37.15 (36.13-38.17)}\\
\midrule
4-shot & PathFiT Disable & 43.26 (42.5-44.02) & 89.71 (89.48-89.93)	 & 43.76 (42.74-44.78) \\
      & PathFiT Enable & \textbf{47.52 (46.67-48.37)} & \textbf{91.78 (91.33-92.23)} & \textbf{49.45 (47.40-51.50)} \\
\midrule
8-shot & PathFiT Disable & 54.30 (53.60-55.00) & 93.50 (93.41-93.60) & 54.47 (53.99-54.95)\\
      & PathFiT Enable & \textbf{54.34 (51.67-57.01)} & \textbf{93.85 (93.08-94.62)} & \textbf{55.72 (53.50-57.94)}\\
\midrule
16-shot & PathFiT Disable & \textbf{63.39 (63.17-63.62)} & \textbf{95.97 (95.92-96.02)} & \textbf{64.40 (64.19-64.62)}\\
      & PathFiT Enable & 59.31 (57.81-60.81) & 95.30 (94.86-95.74) & 61.89 (60.31-63.48)\\
\bottomrule
\end{tabular}
\caption{\textbf{Few-shot results of UNI using text prompt learning on pan-cancer classification (TCGA)} in terms of balanced accuracy, ROC AUC, and weighted F1 score. The best-performing model for each metric is bolded, with 95\% CI in parentheses.}
\label{tab:few-shot-tcga-uni}
\end{table}

\begin{table}
\centering
\begin{tabular}{lllll}
\toprule
Shot number &  Adaptation &  Balanced accuracy &  ROC AUC    & Weighted F1\\
\midrule
1-shot & PathFiT Disable & 55.84 (51.53-60.15) & 90.22 (88.52-91.93) & 49.98 (37.95-62.02)\\
      & PathFiT Enable & \textbf{63.92 (60.75-67.1)} & \textbf{94.44 (93.27-95.61)}	 & \textbf{59.31 (50.59-68.04)}\\
\midrule
2-shot & PathFiT Disable & 72.03 (70.27-73.80) & 95.35 (94.69-96.01)	 & 68.76 (63.53-73.98)\\
      & PathFiT Enable & \textbf{76.35 (74.65-78.05)} & \textbf{96.81 (96.49-97.13)} & \textbf{69.10 (64.47-73.72)}\\
\midrule
4-shot & PathFiT Disable & 82.20 (80.60-83.80) & 97.42 (96.93-97.92) & 79.30 (74.40-84.19)\\
      & PathFiT Enable & \textbf{85.97 (82.52-89.42)} & \textbf{98.46 (98.07-98.84)} & \textbf{83.34 (79.57-87.10)}\\
\midrule
8-shot & PathFiT Disable & 90.24 (89.55-90.94) & 99.07 (98.88-99.26) & 88.10 (86.78-89.43)\\
      & PathFiT Enable & \textbf{91.23 (90.29-92.17)} & \textbf{99.37 (99.20-99.53)} & \textbf{89.54 (87.90-91.17)}\\
\midrule
16-shot & PathFiT Disable & 93.21 (92.85-93.56) & 99.45 (99.40-99.49) & 89.74 (89.15-90.33) \\
      & PathFiT Enable & \textbf{93.63 (93.26-94.00)} & \textbf{99.49 (99.40-99.59)}	 & \textbf{91.05 (89.82-92.27)}\\
\bottomrule
\end{tabular}
\caption{\textbf{Few-shot results of CONCH using text prompt learning on ESCA tissue classification (TolkachData)} in terms of balanced accuracy, ROC AUC, and weighted F1 score. The best-performing model for each metric is bolded, with 95\% CI in parentheses.}
\label{tab:few-shot-esca-conch}
\end{table}

\begin{table}
\centering
\begin{tabular}{lllll}
\toprule
Shot number &  Adaptation &  Balanced accuracy &  ROC AUC    & Weighted F1\\
\midrule
1-shot & PathFiT Disable & 60.35 (58.25-62.45) & 90.21 (89.07-91.36) & 49.91 (45.75-54.07)\\
      & PathFiT Enable & \textbf{69.43 (65.63-73.24)} & \textbf{93.68 (92.33-95.03)} & \textbf{56.04 (44.42-67.65)}\\
\midrule
2-shot & PathFiT Disable & 77.32 (75.53-79.12) & 95.88 (95.42-96.33) & 70.45 (67.06-73.84)\\
      & PathFiT Enable & \textbf{79.50 (76.51-82.49)} & \textbf{96.13 (95.05-97.22)}& \textbf{71.17 (62.39-79.96)}\\
\midrule
4-shot & PathFiT Disable & 87.34 (86.07-88.60) & 98.26 (98.09-98.42)	 & 82.83 (81.16-84.51)\\
      & PathFiT Enable & \textbf{88.95 (86.90-91.00)} & \textbf{98.44 (98.10-98.77)} & \textbf{88.32 (86.24-90.40)}\\
\midrule
8-shot & PathFiT Disable & 91.00 (90.39-91.60) & 99.03 (98.98-99.09) & 87.74 (87.19-88.30)\\
      & PathFiT Enable & \textbf{91.24 (89.78-92.70)} & \textbf{99.12 (98.82-99.42)} & \textbf{88.99 (86.77-91.22)}\\
\midrule
16-shot & PathFiT Disable & 93.65 (93.13-94.17) & \textbf{99.46 (99.40-99.52)} & 90.25 (89.48-91.02) \\
      & PathFiT Enable & \textbf{94.46 (94.01-94.91)} & 99.40 (99.32-99.47) & \textbf{92.40 (91.63-93.17)}\\
\bottomrule
\end{tabular}
\caption{\textbf{Few-shot results of UNI using text prompt learning on ESCA tissue classification (TolkachData)} in terms of balanced accuracy, ROC AUC, and weighted F1 score. The best-performing model for each metric is bolded, with 95\% CI in parentheses.}
\label{tab:few-shot-esca-uni}
\end{table}

\begin{table}
\centering
\begin{tabular}{lllll}
\toprule
Shot number &  Adaptation &  Balanced accuracy &  ROC AUC    & Weighted F1\\
\midrule
1-shot & PathFiT Disable & 60.54 (54.85-66.24) & 92.59 (91.22-93.96) & 63.45 (57.37-69.53)\\
      & PathFiT Enable & \textbf{65.61 (62.57-68.66)} & \textbf{96.48 (95.69-97.27)} & \textbf{65.16 (57.77-72.55)}\\
\midrule
2-shot & PathFiT Disable & 72.39 (68.35-76.43) & 96.34 (95.28-97.39) & 75.20 (69.95-80.46)\\
      & PathFiT Enable & \textbf{80.55 (78.12-82.99)} & \textbf{97.80 (97.46-98.15)} & \textbf{78.10 (74.93-81.27)}\\
\midrule
4-shot & PathFiT Disable & 82.68 (78.68-86.69) & 98.73 (98.47-98.98) & 84.08 (79.60-88.55)\\
      & PathFiT Enable & \textbf{86.71 (82.76-90.65)} & \textbf{98.86 (98.24-99.49)} & \textbf{88.83 (84.42-93.25)}\\
\midrule
8-shot & PathFiT Disable & 90.67 (89.19-92.15) & \textbf{99.59 (99.54-99.63)} & 92.75 (91.64-93.86)\\
      & PathFiT Enable & \textbf{92.94 (92.07-93.81)} & 99.46 (99.3-99.62)	 & \textbf{94.84 (94.36-95.32)}\\
\midrule
16-shot & PathFiT Disable & \textbf{92.94 (91.9-93.97)} & \textbf{99.68 (99.65-99.71)} & \textbf{95.14 (94.74-95.54)} \\
      & PathFiT Enable & 92.18 (91.20-93.16) & 99.33 (99.07-99.58) & 94.14 (93.09-95.19)\\
\bottomrule
\end{tabular}
\caption{\textbf{Few-shot results of CONCH using text prompt learning on CRC tissue classification (CRC-100K)} in terms of balanced accuracy, ROC AUC, and weighted F1 score. The best-performing model for each metric is bolded, with 95\% CI in parentheses.}
\label{tab:few-shot-crc100k-conch}
\end{table}

\begin{table}
\centering
\begin{tabular}{lllll}
\toprule
Shot number &  Adaptation &  Balanced accuracy &  ROC AUC    & Weighted F1\\
\midrule
1-shot & PathFiT Disable & 60.31 (57.17-63.44) & 92.35 (90.87-93.84) & 64.77 (62.62-66.92)\\
      & PathFiT Enable & \textbf{66.21 (58.99-73.44)} & \textbf{97.17 (96.83-97.51)} & \textbf{71.04 (60.50-81.58)}\\
\midrule
2-shot & PathFiT Disable & 69.37 (67.09-71.65 & 95.02 (94.27-95.77)	 & 74.01 (70.96-77.05)\\
      & PathFiT Enable & \textbf{78.76 (75.35-82.17)} & \textbf{97.58 (96.30-98.87)} & \textbf{82.78 (78.54-87.01)}\\
\midrule
4-shot & PathFiT Disable & 79.25 (77.04-81.46) & 97.91 (97.38-98.44) & 81.91 (79.58-84.25)\\
      & PathFiT Enable & \textbf{88.46 (85.92-91.00)} & \textbf{98.98 (98.43-99.53)} & \textbf{91.45 (89.73-93.16)}\\
\midrule
8-shot & PathFiT Disable & 88.38 (87.32-89.45) & \textbf{99.04 (98.72-99.37) }& 91.11 (90.45-91.76)\\
      & PathFiT Enable & \textbf{89.43 (86.63-92.22)} & \textbf{99.04 (98.61-99.48)} & \textbf{91.72 (88.9-94.54)}\\
\midrule
16-shot & PathFiT Disable & 90.58 (89.96-91.20) & \textbf{99.36 (99.18-99.53)} & 93.28 (92.70-93.85) \\
      & PathFiT Enable & \textbf{92.56 (89.49-95.63)} & 98.70 (97.78-99.63)	 & \textbf{94.81 (93.00-96.63)}\\
\bottomrule
\end{tabular}
\caption{\textbf{Few-shot results of UNI using text prompt learning on CRC tissue classification (CRC-100K)} in terms of balanced accuracy, ROC AUC, and weighted F1 score. The best-performing model for each metric is bolded, with 95\% CI in parentheses.}
\label{tab:few-shot-crc100k-uni}
\end{table}

\clearpage

\begin{nolinenumbers}
\section*{References} 
\vspace{2mm}

\begin{spacing}{0.9}
\bibliographystyle{naturemag}
\bibliography{sample}
\end{spacing}
\end{nolinenumbers}

\end{document}